\documentclass[aps,superscriptaddress,twocolumn,longbibliography]{revtex4-1}
\usepackage{amsmath,bm}
\usepackage{amssymb}
\usepackage{graphicx}
\usepackage{subcaption}
\usepackage{multirow}
\usepackage{slashed}
\usepackage{amsfonts}
\usepackage{verbatim}
\usepackage{appendix}
\usepackage{xcolor}
\usepackage{makecell} 
\usepackage[version=4]{mhchem}
\usepackage{enumerate}

\usepackage{multirow}
\usepackage{mathrsfs}
\usepackage[colorlinks, breaklinks=true,linkcolor=red, citecolor=blue, linktocpage=true]{hyperref}
\usepackage{cleveref}
\usepackage{footmisc}

\usepackage{color}

\def\green{\color{black}}

\newcommand{\magen}{\color{black}}


\begin{document}

	\title{Nonlinear Quantum Electrodynamics in Dirac materials}

	\author{Ayd\i n Cem Keser}
	\affiliation{School of Physics, University of New South Wales, Sydney, NSW 2052, Australia}
	\affiliation{Australian Research Council Centre of Excellence in Low-Energy Electronics Technologies,The University of New South Wales, Sydney 2052, Australia}

	\author{Yuli Lyanda-Geller}
	\affiliation{Department of Physics and Astronomy, Purdue University, West Lafayette, IN, 47907, USA}
	\author{Oleg P. Sushkov}
	\affiliation{School of Physics, University of New South Wales, Sydney, NSW 2052, Australia}
	\affiliation{Australian Research Council Centre of Excellence in Low-Energy Electronics Technologies,The University of New South Wales, Sydney 2052, Australia}

	\date{\today}

	\begin{abstract}
		Classical electromagnetism is linear. However, fields can polarize the vacuum Dirac sea, causing quantum nonlinear electromagnetic phenomena, e.g., scattering and splitting of photons, that occur only in very strong fields found in neutron stars or heavy ion colliders. 
		We show that strong nonlinearity arises in Dirac materials at
		much lower fields $\sim 1\:\text{T}$, allowing us to explore the  nonperturbative, extremely high field limit of quantum electrodynamics in solids.  We explain recent experiments in a unified framework and predict a new class of nonlinear magneto-electric effects, including a magnetic  enhancement of dielectric constant of insulators and a strong electric modulation of magnetization. We propose experiments and discuss the applications in novel materials. 
	\end{abstract}	
	
	\maketitle

	Classical electromagnetism is  linear and hence supports the principle of
	superposition.   It has been pointed out by Heisenberg and Euler in 1936
	that due to quantum  mechanical effects and the presence of the Dirac sea, linearity ceases to hold in
	strong fields~\cite{euler-heisenberg-german,*euler-heisenberg}.
	Quantum electrodynamics (QED) is therefore nonlinear as the electromagnetic field polarizes the Dirac sea as though it is a material medium. This effect becomes  significant at electric and magnetic fields
	$E_\star \simeq 1.3\times 10^{16}\:\text{V}/\text{cm}$,
	$B_\star \simeq 4.4 \times 10^9\: \text{T} $, at which the Zeeman splitting and electric potential over the Compton wavelength become comparable to the electron rest energy.
	These are the so called Schwinger critical values~\cite{schwinger} and
	they are enormous on the laboratory scale.
	Such fields exist only in exotic environments e.g. neutron stars~\cite{Kaspi2017,*reisenegger2001magnetic} and heavy ion
	colliders~\cite{heavyion}.
	Nevertheless, some low-order nonlinear QED effects, such as scattering or splitting of photons  have been observed in the laboratory{\magen
		~\cite{photonphoton,Akhmadaliev1998}, and probing strong field effects is an active area of research~\cite{Fedeli2021}.
	}
	
	Dirac materials have been known for decades~\cite{Cohen1960, Lax1960, Wolff1964}.
	Nevertheless, their recently understood topological properties  and surface excitations have led to a surge of interest
	~\cite{Qi2008,Zhang2009,Hasan2010,Ando2013,Wiehling2014,Culcer2020}. {  The nonlinear electromagnetic response of Dirac materials have been studied ~\cite{Sodemann2015,Matsyshyn2019,Culcer2020,rostami2020dominant,rostami2020manybody,Bhalla2019,du2020quantum, Morimoto} due to their transport properties (e.g. rectification) and possible applications in photovoltaics}. In this letter, rather than transport, we study dielectric and magnetization response of the three-dimensional (3D) Dirac insulators and semimetals due to the Dirac vacuum, i.e. filled valence band. We include nonlinear contributions to all orders by nonperturbatively analyzing the Heisenberg-Euler
	action~\cite{euler-heisenberg,Akhiezer1965,Berestetskii1982,Peskin2005, Dunne,*Dunne_arxiv}, going both beyond known results of QED and the general framework in condensed matter physics~\footnote{The high order nonlinear electromagnetic effects can also emerge in two-dimensional materials~\cite{Katsnelson2013}, but is unrelated to the effects we discuss here, due to dimensionality considerations. For applications to SU(N) gauge fields see Ref.~\cite{SU(N)}}. 
	\begin{figure}
		\centering
		\includegraphics[width=.7\columnwidth]{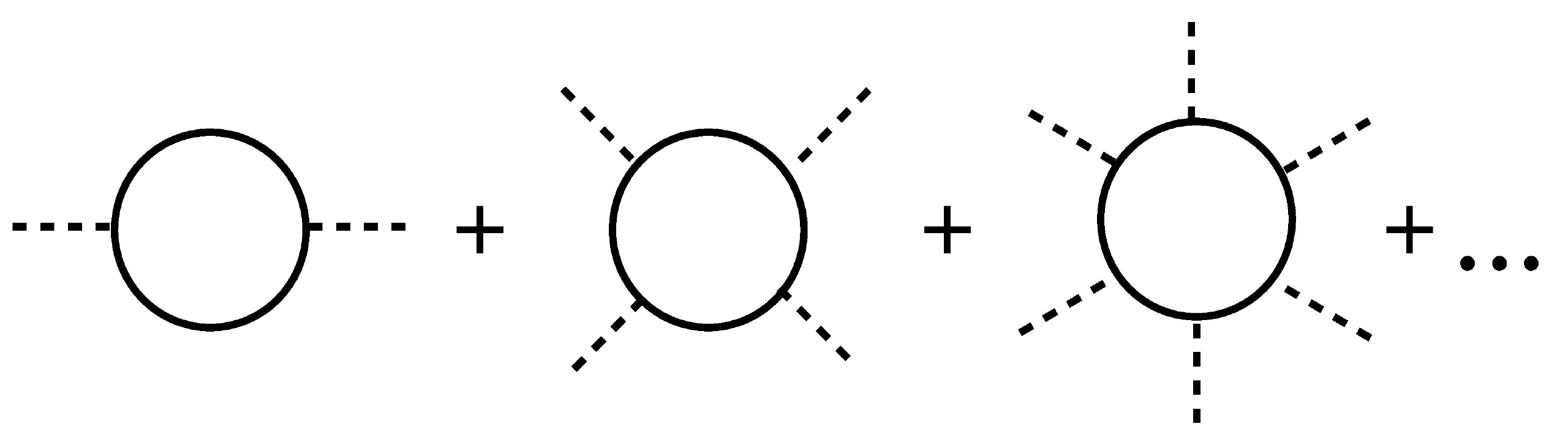}
		\caption{Diagrammatic representation of the Heisenberg-Euler action $\delta L_1 + \delta L_{HE}$ (Eqs.~\eqref{dL1} and \eqref{Schwinger}).
			The dashed line corresponds to the constant external electromagnetic
			fields $B$,$E$ and the solid line is the Green function of a Dirac sea electron.
		}
		\label{HEchain}
	\end{figure}
	In known Dirac materials,  we find the typical values  of Schwinger fields, 	
	$E_\star \sim  10^{5}\:\text{V}/\text{cm}$, $B_\star \sim 1\: \text{T} $,
	are easily accessible,  providing a platform to explore the strong field regime of QED and to observe quantum nonlinear electromagnetic effects in the laboratory.

	The nonlinear effects contribute to the experimentally observed high-field magnetization in the recent work on the Weyl semimetal TaAs~\cite{Zhang2019} and the Dirac semimetal Bi~ \cite{Iwasa2019}, but the importance of this observation and its  origin in the Heisenberg-Euler effect has not been recognized. In the present work we demonstrate this connection and show that the data ~\cite{Zhang2019,Iwasa2019} agree with our predictions. More importantly, we predict  a new class of magneto-electric effects.
	The most significant is the magnetic field tunable, very large enhancement of the dielectric constant, reaching up to $\delta\epsilon_r  \sim 10$ per every $1\:\text{T}$ of the applied magnetic field.
	We also predict an electric field modulated  magnetization.
	Both these effects are highly anisotropic, that is, they depend on relative orientation of E and B-fields and their crystallographic orientation.

	In a material, the classical  Lagrangian of the electromagnetic field is~\footnote{In CGS units.  In an anisotropic material $\bm{\epsilon},\bm{\mu}$ are symmetric $3\times3$ matrices. Some QED textbooks are opposite to the condensed matter convention, see the footnote in ${\S}$ 129 of Ref.~\cite{Berestetskii1982}. In this paper $\bm{B} = \bm{\mu} \bm{H}$ is the magnetic flux density and, $\bm{E} = \bm{\epsilon}^{-1}\bm D$ is the screened electric field.}
	\begin{eqnarray}
		\label{Lcl}
		L_{cl}=\frac{1}{8\pi	}( \bm{E}^T \bm{\epsilon} \bm{E} - \bm{B}^T \bm{\mu}^{-1} \bm{B}).
	\end{eqnarray}
	One of our principal results is the quantum 1-loop, nonlinear,  nonperturbative contribution to the Lagrangian 
	\begin{equation}
		\label{strong}
		\delta L_{HE}
		\to \frac{\Delta}{24\pi^2 \lambdabar_D^3}\left[({\bf b}\cdot{\bf e})^2
		|{\bf b}|^{-1}
		+|\mathbf{b}|^2\ln |{\bf b}| \right],
	\end{equation}
	in a strong B, and weak E-field { (See further below and also  Sec.~\ref{sec:asymptotic} of the supplement}). Here, the dimensionless {\green {vectors $\mathbf{e}$ and $\mathbf{b}$ depend on the fine structure constant $\alpha_D=e^2/\hbar v$ of the Dirac material}}:
	\begin{equation}
		\label{eb}
		{\bf e}(\alpha_D)= \frac{\bm{\mathcal{U} E}}{ E_\star(\alpha_D)}, \quad   {\bf b}(\alpha_D)= \frac{ \bm{\mathcal{U}^{-1} B}}{B_\star (\alpha_D)},
	\end{equation} 
	and the critical `Schwinger' electric $E_\star(\alpha_D)$ and magnetic $B_\star(\alpha_	D)$  fields in the material are defined by 
	\begin{equation}
		\label{eb_crit}	
		E_\star^2(\alpha_D) =\frac{v^2}{c^2 }B_\star^2(\alpha_D) =\frac{\Delta}{\alpha_D \lambdabar_D^3} =\left(\frac{\Delta^2}{e\hbar v}\right)^2.
	\end{equation} 
	Eqs.~\eqref{strong}, ~\eqref{eb} and \eqref{eb_crit}  account for the anisotropy of real materials~\cite{Aronov}, for which the velocity tensor is a $3\times3$ symmetric matrix~\cite{Wolff1964} $\bm{\mathcal{V}} = v \bm{\mathcal{U}}$, with $\text{det}(\bm{\mathcal{U}})=1$. The term $\bm{\mathcal{U} E}$ and $\bm{\mathcal{U}^{-1} B}$, are linear transformations of $\bm{E}$ and $\bm{B}$ respectively (See
	Supplement Sec.~\ref{s-sec:aniso}~\cite{Supplementary_material}).
	
	\begin{table}[t]
		\begin{tabular}{lllll}
			\hline
			& \multirow{1}{*}{$2\Delta$ [meV]} & \multirow{1}{*}{${\alpha_D}/{\alpha}$} &\multirow{1}{*}{ $E_\star$ [V/cm]} &\multirow{1}{*}{ $B_\star$ [mT]} 		\vspace{2pt} \\ 
			\hline
			QED  $(\Delta = m_e c^2)$   &   $10^{9}$ ~\cite{Thomson,*Millikan,CODATA}    & 1         &    $1.3\times 10^{16}$     &  $4.4\times 10^{12}$   \\
			$\text{Pb}_{0.5}\text{Sn}_{0.5}\text{Te}$      &   $63$ ~\cite{Dziawa2012}     & 580	         &    $2.9\times 10^4$     &  $5.6\times 10^3$   \\
			$\text{Bi}_{0.9}\text{Sb}_{0.1}$ &  $15.5$  ~\cite{Liu1995,Hsieh2008}     &     188     &  571      &  36  \\ 
			TaAs    &    $0$    &     357   &     0   &      0 \\ 
		\end{tabular}
		\caption{Comparison of parameters including the band gap ($2\Delta$)~\cite{edotb}, effective fine structure constant  $\alpha_D = \frac{e^2}{\hbar v}$ as the ratio $\alpha_D/\alpha = c/v$, and the Schwinger fields in Eq.~\eqref{eb_crit}.  }
		\label{tab:materials}
	\end{table}
	
	We have defined the symbols in Eqs.(\ref{eb}),(\ref{eb_crit})
	according to convention in QED.
	The ``Dirac wavelength'' $\lambdabar_D=\frac{\hbar v}{\Delta}$ and the   ``Dirac magneton'' $\mu_D=\frac{e\hbar v^2}{2\Delta c}$ replace the Compton wavelength and the Bohr magneton respectively.  When the fields reach the `Schwinger scale',
	Zeeman splitting and the potential difference at $\lambdabar_D$ are equal to the half of the Dirac  band-gap:
	\begin{eqnarray}
		\label{Sf}
		&&2\mu_DB_\star=\lambdabar_DeE_\star=\Delta,
	\end{eqnarray}
	and the nonlinearity becomes relevant. 
	In  Table~\ref{tab:materials} we list material parameters considered in this work. For more details see
	Sec.~\ref{sec:QED_compare} and Table~\ref{table:compare} of the Supplement \cite{Supplementary_material}.

	\begin{figure*}[t]
		\centering
		\begin{subfigure}{0.3\textwidth}
			\includegraphics[width=
			\textwidth]{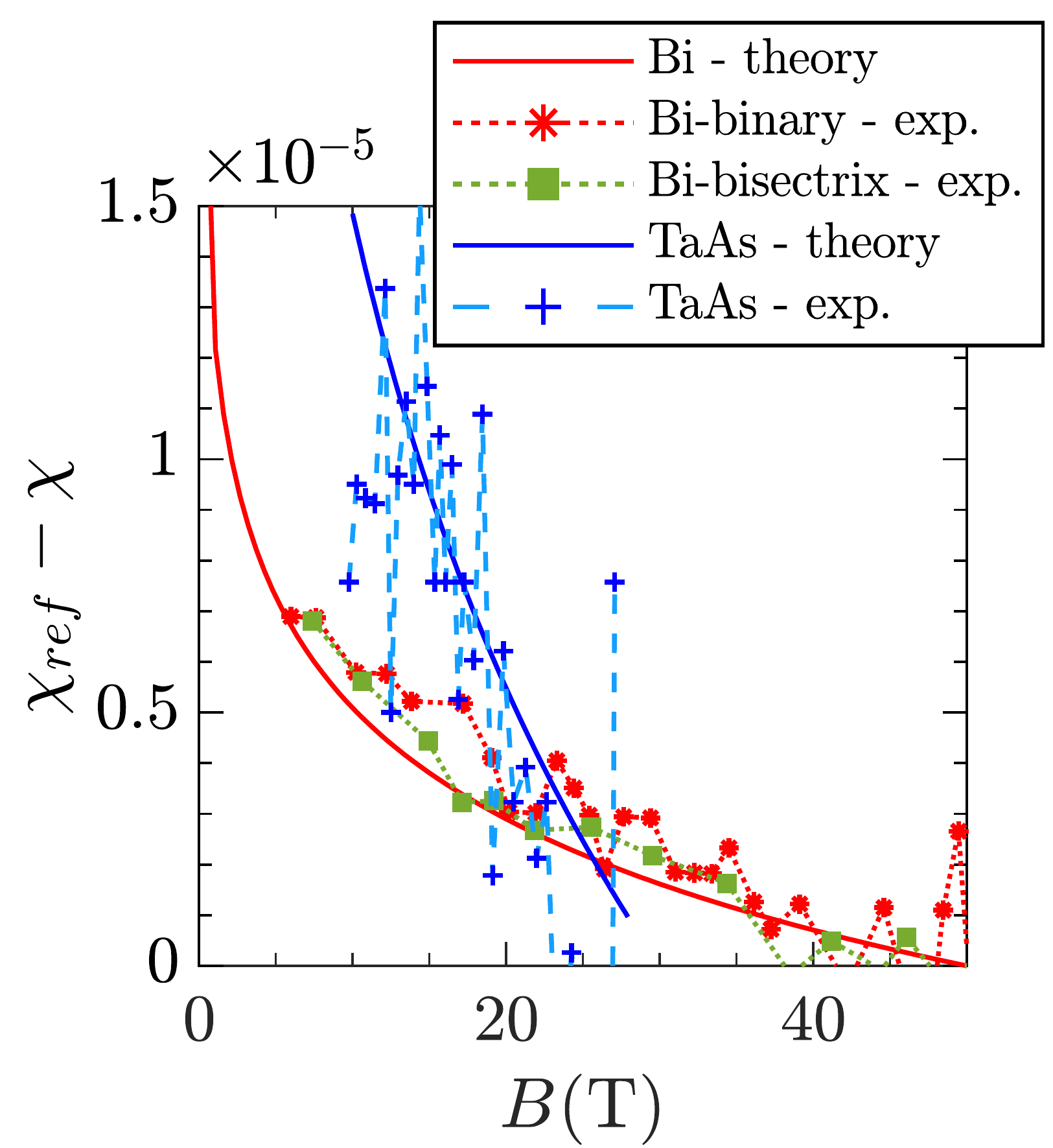}
			\caption{}
			\label{fig:chis}
		\end{subfigure}
		\begin{subfigure}{0.3 	\textwidth}
			\includegraphics[width=
			\textwidth]{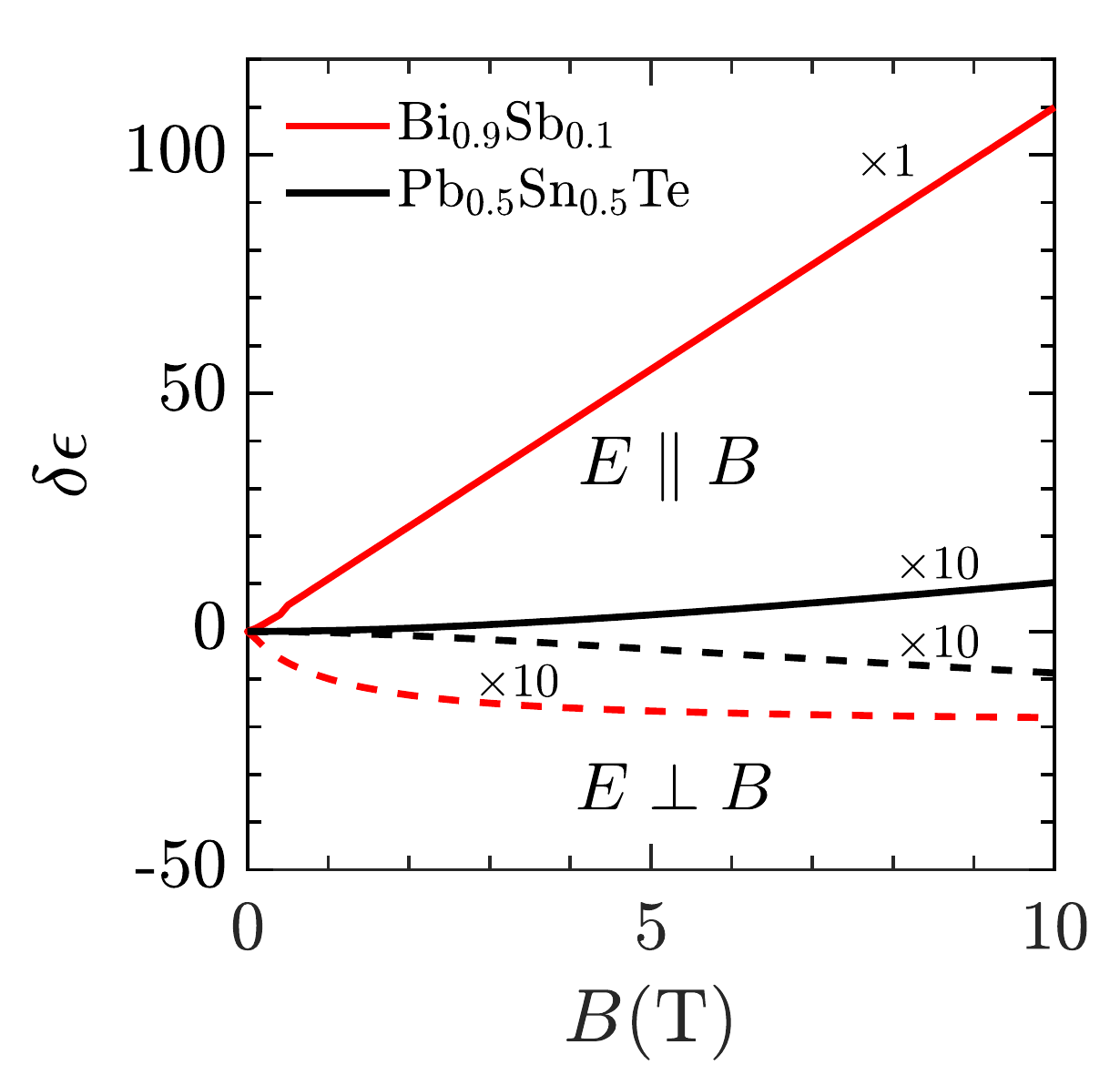}
			\caption{}
			\label{fig:deps}
		\end{subfigure}
		\begin{subfigure}{0.3 \textwidth}
			\includegraphics[width=
			\textwidth]{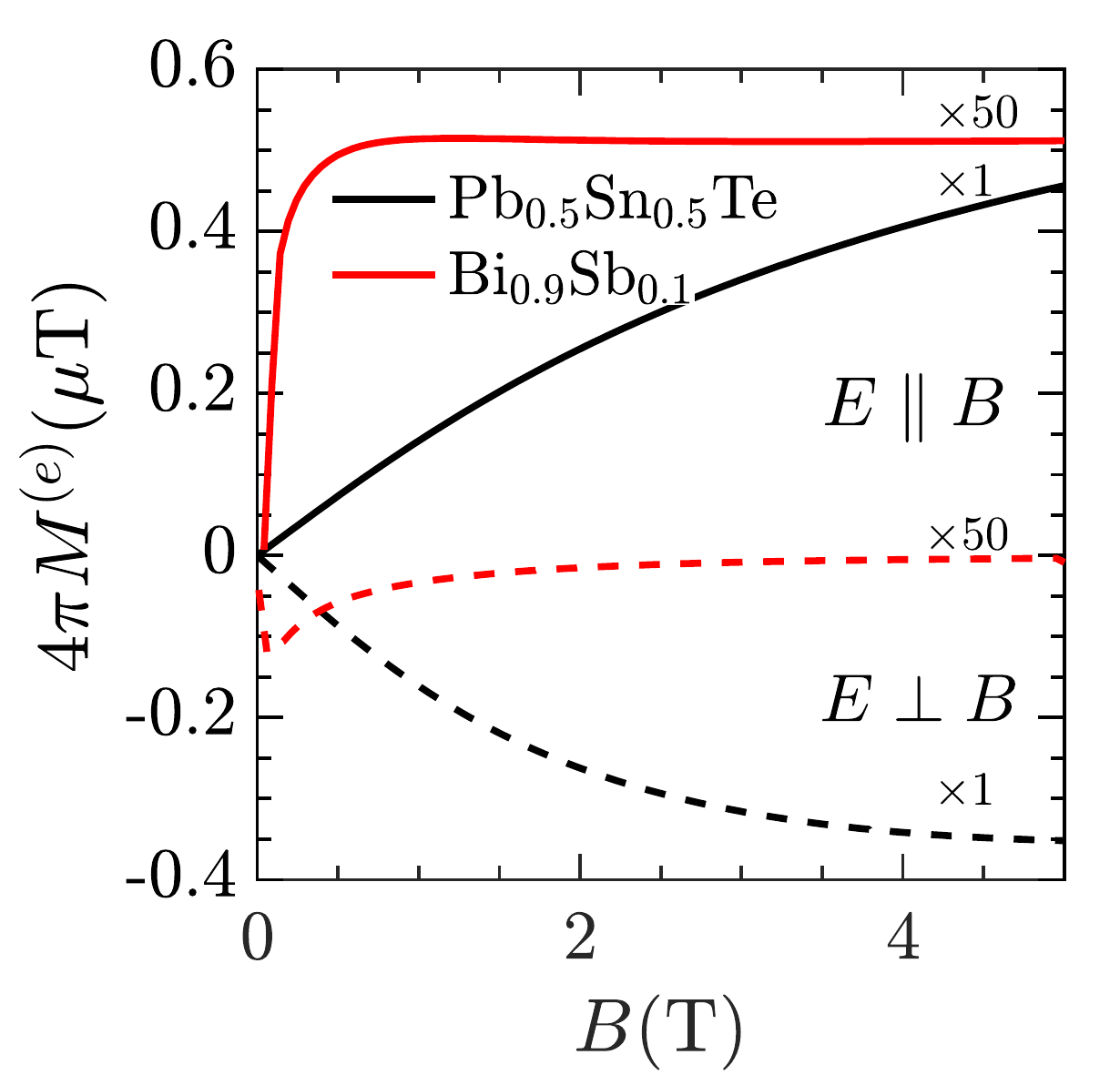}
			\caption{}
			\label{fig:magn}
		\end{subfigure}
		\caption{\magen
			a) Nonlinear diamagnetic susceptibility $\chi_{ref} -\chi$ versus magnetic field,
			$\chi_{ref} = \chi(50\:\text{T})$ in Bi and 	$\chi_{ref} = \chi(30\:\text{T})$ in 
			TaAs. In TaAs (blue) the field is along the crystal c-direction and in Bi there are two directions, binary (red)  and bisectrix (green). 
			The points represent numerical differentiation
			of TaAs and Bi magnetization data from Refs.~\cite{Zhang2019} and ~\cite{Iwasa2019} respectively. The experimental points are connected 	by dashed lines for guidance. Solid lines represent our theory.
			The red solid line is our prediction for $\text{Bi}_{0.9}\text{Sb}_{0.1}$.
			b) Predicted variation of the dielectric constant in magnetic field, for parallel (perpendicular) field configurations shown by solid (dashed) lines.
			c) Predicted electric field modulated magnetization as a function of applied magnetic field, along the magnetic field direction at $E= 0.3 E_\star$.}
		\label{fig:epsilon_M}
	\end{figure*}

	The quantum contribution to the Lagrangian can be viewed as the sum of the infinite chain of 1-loop diagrams in Fig.\ref{HEchain} that represent the polarization of the Dirac sea of electrons by external electric and magnetic fields.  In this work we consider only non-magnetic crystals with inversion symmetry~\cite{Fu2007,*Bansil2016}~\footnote{TaAs is non-centrosymmetric crystal, but since it is gapless, we take $E=0$ to avoid transport. TaAs is non-magnetic and the time reversal symmetry requires that the effective action contains even powers of B only.} { and assume} the static/quasistatic approximation,  $\omega, kv\ll \Delta$, where $\omega$ and $k$ are the frequency and the wave number of the external fields.
	Therefore our diagrams, Fig.\ref{HEchain},
	have only even numbers of external E-lines and B-lines.
	Besides diagrams in Fig.\ref{HEchain}, there are also multi-loop diagrams suppressed by a factor of  $\alpha_D/\epsilon \sim 0.03$ per each additional loop, 
	where $\epsilon$ is the large dielectric constant mainly due to the lattice and intra-ionic polarization. For the discussion of the suppression of the multi-loop diagrams in the context of phenomena considered here,
	see Sec.~\ref{sec:eff_act} in the supplement
	and also Refs.~\cite{Sham1966,Sham_Rice1966}

	In Fig.~\ref{HEchain}, the first diagram quadratic in external fields is ultraviolet divergent and is equal to~\cite{Akhiezer1965,edotb}
	\begin{equation}
		\label{dL1}
		\delta L_1 =
		\frac{\Delta }{12\pi^2\lambdabar_D^3}\ln\left(\frac{\Lambda}{\Delta}\right) \left( |\bf{e}|^2 -   |\bf{b}|^2\right).
	\end{equation} 
	Here the subscript `1' indicates contribution from the first diagram in Fig.\ref{HEchain} and
	$\Lambda \sim v\frac{\hbar\pi}{a} \sim 1\:\text{eV}$ is
	the ultraviolet cut-off energy, where $a$ being the lattice spacing.
	In QED this diagram describes the electric permittivity and magnetic permeability of vacuum and thus it is included 
	into the definitions of the electric charge and electromagnetic fields.
	As a result, $\delta L_1 $ does not appear explicitly in QED. 
	However, for Dirac
	materials $\delta L_1 $ is an explicit contribution that has to be added
	to the classical Lagrangian Eq.~\eqref{Lcl}.
	Indeed, this is the contribution of the Dirac sea (valence band) to the dielectric constant and magnetic susceptibility.

	Equating $(E^2-B^2)/(8\pi)+\delta L_1$ to the classical Lagrangian (\ref{Lcl}), we find the linear dielectric constant  $\bm{\epsilon_D}$ and the linear magnetic susceptibility $\bm{\chi_D}$ ($\bm{\mu} = \bm{1} + 4\pi\bm{\chi}$):
	\begin{eqnarray}
		\label{ec}	&&\bm{\epsilon_D}=\bm{1}+\frac{2\alpha_D}{3\pi} \ln\left(\frac{\Lambda}{\Delta}\right)\bm{\mathcal{U}}^2, \quad \epsilon_D
		\sim 3 , \\
		\label{ec2}
		\label{mag}&&\bm{\chi_D}=-\frac{\alpha_D}{6\pi^2}\frac{v^2}{c^2}
		\ln\left(\frac{\Lambda}{\Delta}\right)\bm{\mathcal{U}}^{-2},\quad \chi_D \sim -10^{-6}.
	\end{eqnarray}
	where estimates are given for the diagonalized tensors.
	
	Eqs.~\eqref{ec} and \eqref{ec2} { define the Dirac contributions
		to the total dielectric and magnetic susceptibilities}.
	The	contribution (\ref{ec}) is  relatively small  compared
	to the total relative permittivity  $\bm{\epsilon}$ in  Eq.~\eqref{Lcl}, typically $\epsilon \sim 100$, which is primarily  due to the ionic (lattice) and intra-ionic contributions (See Supplement Table~\ref{table:epsilons}).
	The magnetic response (\ref{mag}) constitutes a significant part  of the diamagnetic susceptibility, which also has contributions from lower bands and core electrons.
	For Bismuth, the Dirac valence band contribution (\ref{mag}) has been previously
	considered in Ref.~\cite{Fuseya2015}.

	We describe now the nonlinear effects.  The diagrams in Fig.\ref{HEchain}
	
	beyond the first one  ($n\geq 2$) are convergent at arbitrarily large $\mathbf{|e|,|b|}$~\footnote{$\mathbf{|e|,|b|} \ll \Lambda^2/\Delta^2$, otherwise the effective Dirac Hamiltonian no longer applies. See Sec.~\ref{sec:validity}} and are re-summed exactly~\footnote{The exact result is obtained by directly summing the energies of Landau levels ~\cite{Akhiezer1965,Berestetskii1982}, or in modern language, zeta-function regularization~\cite{Dunne,*Dunne_arxiv}}  to yield the 1-loop,  nonperturbative Heisenberg-Euler action
	\begin{align}
		\label{Schwinger}
		\delta L_{HE}&= \sum_{n=2}^{\infty} \delta L_n \equiv 
		\frac{-\Delta}{8\pi^2\lambdabar_D^3} \int_0^\infty \frac{d\eta e^{-\eta}}{\eta}
		\nonumber\\
		\times\bigg[A_- &\cot (\eta A_-) A_+ \cot(\eta A_+)
		- \frac{1}{\eta^2} + \frac{1}{3} (A_-^2+A_+^2)\bigg],\nonumber \\
		A_\mp&=-\frac{i}{2}\left[\sqrt{({\bf b}+i{\bf e})^2}
		\mp\sqrt{({\bf b}-i{\bf e})^2}\right],
	\end{align} 
	which  accounts for crystal anisotropy, cf.,  Eq.~\eqref{eb}, as well as the strong field behavior.  The imaginary part of Eq.(\ref{Schwinger}), obtained via its analytic continuation, captures the electric breakdown, which can be avoided in weak electric fields $|{\bf e}| < 1$ ($E < E_*$ ). Then, Eq.~\eqref{Schwinger} can be expanded in powers of $\mathbf{e}$. However, the magnetic field can be much larger than $B_\star$, leading to the asymptotic expression  Eq.~\eqref{strong} (See Supplement Sec.~\ref{sec:asymptotic}). At weak magnetic fields, $|{\bf e}|,|{\bf b}|\ll 1$,  Eq.~\eqref{Schwinger}  reduces to the 2nd diagram in Fig.\ref{HEchain},
	\begin{eqnarray}
		\label{weak}
		\delta L_2=
		\frac{\Delta}{360\pi^2\lambdabar_D^3}
		\left[\left(|\mathbf{e}|^2- |\mathbf{b}|^2\right)^2
		+7\left({\bf e}\cdot{\bf b}\right)^2\right].
	\end{eqnarray}

	At $E=0$,  the nonlinear magnetic
	susceptibility is
	\begin{eqnarray}
		\label{dchi4}
		&&
		\bm{\delta\chi}=\frac{\partial^2\delta L_{HE}}{\partial \bm{B}\partial \bm{B}}
		=\bm{\mathcal{U}}^{-2}\frac{\alpha_D}{12\pi^2}\frac{v^2}{c^2}F(|{\bf b}|);
		\nonumber\\
		&&  F(|{\bf b}|)=\frac{2}{5}|{\bf b}|^2,  \ \ \|{\bf b}|\ll 1\nonumber\\
		&& F(|{\bf b}|)= \ln |{\bf b}|, \ \ \ |{\bf b}|\gg 1.
	\end{eqnarray}
	The dimensionless function $F(|{\bf b}|)$ in the full range of magnetic fields
	obtained by numerical integration of Eq.~\eqref{Schwinger}
	is shown  in  Supplement Fig.~\ref{fig:FGD}. 
	Strong and weak field limits of $F$ follow from the actions given by
	Eqs.(\ref{strong}) and (\ref{weak}) respectively.
	
	The total magnetic susceptibility of the Dirac valence band is the sum of the linear susceptibility,
	Eq.(\ref{mag}) and the nonlinear  contribution, $\bm{\chi}=\bm{\chi_D}+\bm{\delta\chi}$.	
	%
	When $|{\bf b}|\gg 1$ we have
	\begin{equation}
		\label{gapless}
		\bm{ \chi}= -\bm{\mathcal{U}}^{-2}\frac{\alpha_D}{12\pi^2}\frac{v^2}{c^2}
		\ln\left(\frac{c\Lambda^2}{e|\bm{\mathcal{U}^{-1} B}|\hbar v^2}\right).
	\end{equation} 	Here $\bm{\chi}$ depends on $B$ but not on $\Delta$, and is well-defined in the limit $\Delta=0$, as in the Weyl semimetal TaAs~\cite{Zhang2019}.

	According to Eqs.(\ref{dchi4}),(\ref{gapless}) the magnetic susceptibility is nonlinear, i.e. it depends on magnetic field.
	Remarkably, this Dirac nonlinearity  has been  recently observed,  but its connection to nonlinear electrodynamics was not identified. 
	Here we show its origin in the Heisenberg-Euler effect.
	The magnetization of Weyl semimetal TaAs has been measured up to
	$B=30$ T, Ref.~\cite{Zhang2019}, and magnetization of  Dirac semimetal Bi
	has been measured up to $B=60$ T, Ref.~\cite{Iwasa2019}. 
	In Zhang  \textit{et al.}~\cite{Zhang2019} the valence band contribution to magnetization at $\bm E=0$ was considered \footnote{ However, our approach from the vantage point afforded by the transformation of Heisenberg-Euler action under dilation and contraction of space (See Sec.~\ref{sec:aniso}) leads to a more accurate overall factor.}, and  in the high magnetic field limit, the  magnetization quasi-linear in the applied B-field is investigated. Here we study the universal {\it nonlinear susceptibility} and eliminate all uncertainties such as the choice of ultraviolet cut-off $\Lambda$, subleading terms and contributions from other bands or core electrons.

	Both TaAs and Bi have nonzero chemical potential and hence have conduction electrons. Therefore at weak magnetic fields both compounds show
	magnetic oscillations. The conduction electrons freeze and the oscillations disappear at $B>5$T in Bi~\cite{Iwasa2019}
	and $B >10-13$T in TaAs~\cite{Zhang2019}.  In these ranges of B, we can compare the data with our predictions. In Fig.~\ref{fig:chis}
	the points show magnetic susceptibilities of TaAs (c-direction) and Bi (binary and bisector directions). The points have significant spread as they are obtained by numerical differentiation of experimental magnetizations from Refs.~\cite{Zhang2019,Iwasa2019}.
	To focus on the nonlinearity, we plot $\chi_{ref}-\chi$, where
	$\chi_{ref}=\chi(B=30T)$ for TaAs and $\chi_{ref}=\chi(B=50T)$ for Bi.
	Solid curves present our theoretical predictions, which is manifestly consistent with the data.
	For discussion of material specific details, anisotropy etc., see Supplement Sec.~\ref{sec:applications}. 
	Interestingly, 	$\text{Bi}_{0.9}\text{Sb}_{0.1}$ 
	alloy has the band structure very close to that of Bi, but with no conduction electrons~\cite{Liu1995,Hsieh2008}, and could be an ideal test platform for our theory.
	The susceptibility of this compound has not been measured yet, and the solid red curve in  Fig.~\ref{fig:chis} shows our theoretical prediction.

	We now consider novel magneto-electric effects.
	The nonlinear dielectric constant is
	\begin{eqnarray}
		\label{deps}
		&&\bm{\delta\epsilon_D}=4\pi\frac{\partial^2\delta L_{HE}}{\partial \bm{E}\partial \bm{E}}
		=\bm{\mathcal{U}}^2\frac{\alpha_D}{3\pi} G_i(|{\bf b}|);\\
		&&  |{\bf b}|\ll 1: \ \ 
		G_{||}(|{\bf b}|)= \frac{1}{3}|{\bf b}|^2 \ , \ \ \ \ 
		G_{\perp}(|{\bf b}|)= -\frac{2}{15}|{\bf b}|^2 	\nonumber\\
		&&|{\bf b}|\gg 1: \ \  G_{||}(|{\bf b}|)=|{\bf b}|
		\ , \ \ \ \ \ \ \ G_{\perp}(|{\bf b}|)=-\ln(|{\bf b}|).\nonumber
	\end{eqnarray}
	Here the index $i=||,\perp$ shows the relative orientation of $\mathbf{e},\mathbf{b}$~\cite{parallel}. Dimensionless functions $G_i(|{\bf b}|)$ in the whole range of $\bf{b}$
	obtained by numerical integration of (\ref{Schwinger})
	are plotted in Supplement Fig.~\ref{fig:FGD}.
	Its strong and weak field limits define actions given by
	Eqs.(\ref{strong}) and (\ref{weak}) respectively.
	The dependence of the dielectric constant on the applied magnetic field is a novel magneto-electric effect. For ${\mathbf{b}} \parallel {\mathbf{e}}$ the contribution $\delta\epsilon_D $ is positive
	and can be very large, while  for ${\mathbf{b}}\perp{\mathbf{e}}$ the contribution
	$\delta\epsilon_D$ is negative. The expressions for  arbitrary angle between ${\mathbf{b}},  {\mathbf{e}}$, and the relation to the angle between applied fields $\bm{B}$ and $\bm{E}$, which is generally different due to properties of the anisotropy transformation are given in Sec.~\ref{sec:suscep}. Furthermore, according to (\ref{Lcl}), (\ref{weak}) there is a nonlinear contribution quadratic in the electric field,
	\begin{equation}
		\label{deps_e}
		\bm{\delta \epsilon_{D}}(\bm{E})  = \bm{\mathcal{U}}^2  \frac{2\alpha_D}{15 \pi} |\mathbf{e}|^2 , \quad |\mathbf{b}| =0,
	\end{equation}
	which is suppressed by $|\mathbf{e}|^2/|\mathbf{b}|$ when $|\mathbf{b}|\gg1$. Notably, at $|\mathbf{e}|,|\mathbf{b}|\ll1$, contributions (\ref{deps}) and (\ref{deps_e}) add up.

	The magnetic field induced variation of the dielectric constant in Eq.(\ref{deps})
	scales as	$\delta\epsilon_D \propto 1/B_\star \propto \Delta^{-2}$.
	Thus, the effect is most significant in a small band-gap Dirac insulators.
	In Fig.~\ref{fig:deps} we plot our predictions for $\text{Bi}_{0.9}\text{Sb}_{0.1}$. For $\mathbf{e} || \mathbf{b}$ the effect is enormous,  $\delta\epsilon_D \sim 10$/Tesla.
	For $\bm{E} \perp \bm{B}$ the effect is smaller and has the negative sign.
	In the same Fig.~\ref{fig:deps} we also plot predictions for
	$\delta\epsilon_D$ in $\text{Pb}_{0.5}\text{Sn}_{0.5}\text{Te}$. This compound has larger gap and therefore the effect is smaller, but still observable. 
	
	One more novel magneto-electric effect is the 
	dependence	of magnetization, on the applied electric field.	
	The electric field dependent magnetization 
	${\bm M^{(\mathbf{e})}}=\frac{\partial\delta L}{\partial {\bm B}}$, in units of ``Dirac magnetons'' per 	``Dirac volume'', reads
	\begin{eqnarray}
		\label{magn}
		&& 4\pi{\bm M^{(\mathbf{e})}}=\frac{\bm{\mathcal{U}}^{-1}\mathbf{b}}{|\mathbf{b}|}
		\frac{\mu_D}{3\pi\lambdabar_D^3}|{\bf e}|^2D_i(|{\bf b}|)\\
		&&  |{\bf b}|\ll 1: \ \ \
		D_{||}(|{\bf b}|)= \frac{2}{3}|{\bf b}|  \ , \ \ \
		D_{\perp}(|{\bf b}|)= -\frac{4}{15}|{\bf b}|    \nonumber\\
		&&|{\bf b}|\gg 1: \ \ \ D_{||}(|{\bf b}|)=1  \ , \ \ \ \ \ \ \
		D_{\perp}(|{\bf b}|)= -\frac{1}{|{\bf b}|}. \nonumber
	\end{eqnarray}
	The direction of the magnetization (\ref{magn}) in a Dirac crystal is defined by the vector ${\bf b}$ and depends on crystal anisotropy as described by Eq. (\ref{eb}). 
	Dimensionless functions $D_i(|{\bf b}|)$ in the whole range of $\bf{b}$
	obtained by numerical integration of (\ref{Schwinger})
	are plotted  in  Fig.\ref{fig:FGD} in Supplementary material.
	For ${\mathbf{b}}  \parallel  {\mathbf{e}}$ the magnetization is large and 
	paramagnetic, while
	for ${\mathbf{b}}   \perp  {\mathbf{e}}$ the magnetization is diamagnetic~\cite{parallel}. 
	Magnetization  (\ref{magn}) is quadratic in the applied electric field	and  as a function of magnetic field, saturates when  $|{\bf b}| \gg 1$.

	To enhance the magnetization in Eq.~\eqref{magn}
	one  needs the electric field as strong as
	possible. However, the field is limited by the dielectric strength, $E_d$ of the material, beyond which dielectric breakdown occurs. 
	The breakdown probability (rate of Zener tunneling by electric field per unit volume) is obtained from Eq.~\eqref{Schwinger}~\cite{Berestetskii1982} and found to be $P\propto |{\mathbf{ e}}|^2e^{-\pi/|{\mathbf{e}}|}$ (See Sec.~\ref{sec:1-loop}).
	The most important here is the exponential dependence, which universally applies to both the Dirac spectrum and quadratic dispersion.  Thus, one expects that $E_d$, is proportional to $E_\star$. Taking two band
	insulators, diamond ($2\Delta \approx 5.5$eV, $E_d \approx 10^7$ V/cm),
	and silicon ($2\Delta \approx 1.14$ eV, $E_d \approx 3\times 10^5$ V/cm),
	as reference materials, we observe that
	the dielectric strength scales as
	$E_d\propto \Delta^2$. Therefore $E_d$ is a fixed fraction of $E_\star$.
	Significant 	$E$-dependent magnetic { effects} Eq.~\eqref{magn} can then be observed for $|\mathbf{e}|= 0.1\text{-}0.3$~\footnote{Of course, there are other factors like the
		dependence of dielectric strength on impurities, the size of the sample, etc.}.
	Furthermore, as usual in solids, setups with huge built-in electric fields in the insulating regime can be explored \cite{Rediker1, *Rediker2}.

	For a fixed  ${\bf e}=E/E_\star$, the electric field modulated magnetization in Eq.~\ref{magn} obeys 
	$M^{(\mathbf{e})}\propto B_\star \propto \Delta^2$, so materials with large gap are preferable, unlike in the dependence of dielectric constant on magnetic field.
	In Fig.~\ref{fig:magn} we plot the predicted  magnetization for $\text{Pb}_{0.5}\text{Sn}_{0.5}\text{Te}$ versus magnetic field
	at $E=10^4\:\text{V/cm}$, which corresponds to ${\bf e}\approx 0.3$.
	For the both fields, ${\bf e}$ and ${\bf b}$, parallel to the c-axis,
	the electric field driven magnetization is $4\pi M^{(\mathbf{e})}\approx 0.2\:\mu\text{T}$ at $B=1T$.
	When $\bm{E\perp B}$, the magnetization changes sign, see Fig.~\ref{fig:magn}.
	In the same figure we also plot the magnetization in 
	$\text{Bi}_{0.9}\text{Sb}_{0.1}$ for ${\bf e}\approx 0.3$.
	Here the effect is smaller due to the smaller Dirac gap.
	
	The electric field driven magnetization
	in $\text{Bi}_{0.9}\text{Sb}_{0.1}$ ($4\pi M^{(\bf e)} \sim 10^{-8}$T)
	and in $\text{Pb}_{0.5}\text{Sn}_{0.5}\text{Te}$ ($4\pi M^{(\mathbf{e})} \sim 2\times 10^{-7}$T) can be feasibly detected in lock-in experiments, in an applied  electric field having a constant and an AC component (with frequency $\omega$). The induced magnetization is then characterized by contributions modulated at frequencies $\omega$ and $2\omega$. Of course, the condition $\hbar\omega\ll 2\Delta$ is assumed fulfilled. Experiments on observation  of $M^{(\mathbf{e})}$ could also take advantage of the SQUID magnetometry, sensitive to magnetization as low as $10^{-15}\:\text{T}/\sqrt{\text{Hz}}$~\cite{Gramolin2020},
	much lower than the predicted values.

	\textit{In conclusion},
	Based on the Heisenberg-Euler theory of the physical vacuum we
	develop the theory of nonlinear electromagnetic effects in Dirac materials.
	We explain the results of two recent experiments 
	on nonlinear contribution to magnetization of Dirac materials.
	We predict two novel magneto-electric effects
	and discuss possible experiments and materials for their observation.

	\begin{acknowledgments}
		{\it Acknowledgements} We thank M. O'Brien, H. Takagi, A. O. Sushkov,  V. M. Shabaev, J. Seidel, A. R. Hamilton, U. Zuelicke and Y. Ashim for
		useful comments.
		YLG was supported by the U.S. Department of Energy, Office of Basic Energy Sciences, Division of Materials Sciences and Engineering under Award DE-SC0010544. He also acknowledges the Gordon Godfrey bequest
		for the support of his visit to UNSW.
		ACK and OS acknowledge the  support from the Australian Research Council
		Centre of Excellence in Future Low Energy Electronics
		Technologies (CE170100039).	
	\end{acknowledgments}

\bibliographystyle{apsrev4-1}
\bibliography{EH-refs}

\begin{thebibliography}{91}
\bibitem [{edo()}]{edotb}%
  \BibitemOpen
  \href@noop {} {}\bibinfo {note} {When the band gap is inverted, as in
  topological insulators, our analysis applies to $|\Delta|$. The additional
  boundary term in the effective action of the electromagnetic field,
  proportional to $\bm{E}\cdot \bm{B}$ in topological
  insulators~\cite{Frohlich} and Weyl
  semimetals~\cite{Burkov,enriched,Interacting_weyl} does not affect the
  nonlinear response which we investigate in this letter.}\BibitemShut {Stop}%
\bibitem [{par()}]{parallel}%
  \BibitemOpen
  \href@noop {} {}\bibinfo {note} {The angle between $\mathbf{e}$ and
  $\mathbf{b}$ in general is different to the one between applied fields
  $\bm{B}$ and $\bm{E}$. The transformation of the angle is given in Sec.~S1.
  The susceptibilities for arbitrary angle between ${\mathbf{b}},
  {\mathbf{e}}$, are given in Sec.~S6 and S7.}\BibitemShut {Stop}%
\end{thebibliography}

\clearpage
\onecolumngrid

\setcounter{figure}{0}  
\setcounter{page}{1} 

\renewcommand{\thefigure}{S\arabic{figure}}
\renewcommand{\thetable}{S\arabic{table}}
\renewcommand{\theequation}{S\arabic{equation}}
\renewcommand{\thesection}{S\arabic{section}}

\part*{\centering Supplemental Material}
\section{Dirac Cone Anisotropy}
\label{sec:aniso}
Here, for the sake of completeness, we provide a prescription to treat anisotropic crystals, adopted from Ref.~\cite{Aronov}. In real crystals the velocity is
a tensor that is represented by a $3\times 3$ real symmetric matrix $\bm {\mathcal{V}}$, so that the Dirac Hamiltonian reads 
\begin{equation}
	\label{Dirac_aniso}
	\bm{H} = \bm{\beta} \Delta- \bm{1}|e|{\phi} +  \sum_{i,j}\bm{\alpha}_i \mathcal{V}_{ij} (p_j + |e|A_j/c),\quad i,j = 1,2,3.
\end{equation}
where $\bm{\alpha}_i, \bm{\beta}$ are  $4\times 4$ Dirac matrices and $\bm{1}$ is a $4\times4$ identity matrix. We define the Dirac velocity $v$ and the anisotropy matrix $\bm{\mathcal{U}}$ as
\begin{equation}
	v^3 = |\text{det}(\bm{\mathcal{V}})|,\quad 	\bm{\mathcal{U}} = \bm{\mathcal{V}}/v.
\end{equation}
\subsection{Dilation/contractions of space that renders the cone isotropic}
The matrix $\bm{\mathcal{U}}$, being a symmetric matrix with $\text{det}(\bm{\mathcal{U}}) = 1$, without loss of generality, induces a volume preserving transformation on coordinates, that is 
\begin{equation}
	\bm{\tilde{x}} =  \bm{\mathcal{U}}^{-1} \bm{x},\quad  	\bm{\tilde{p}} =  \bm{\mathcal{U}} \bm{p},\quad  	\bm{\tilde{A}} = \bm{\mathcal{U}} \bm{A},\quad d^3 \tilde{x} = d^3 x  {  ,}
\end{equation}
with which the Dirac Hamiltonian assumes its isotropic form
\begin{equation}
	\label{Dirac_iso}
	\bm{H} =   v \bm{{\alpha}}\cdot  (\bm{\tilde{p}} + |e|\bm{\tilde{A}}/c)+ \bm{\beta} \Delta- \bm{1}|e|{\phi}  { .} 
\end{equation}
The Lagrangian, being a scalar, is invariant under this volume preserving  transformation.
Meanwhile,  the scaled electric and magnetic  fields are
\begin{equation}
	\bm{\tilde{E}} = \frac{\partial \phi}{\partial \bm{\tilde{x}}} =  \bm{\mathcal{U} E},\quad  \bm{\tilde{B}} =  \bm{\mathcal{U}^{-1} B} {  ,}
\end{equation} 
where  the equation for $\bm{\tilde{B}}$ follows from the  identity (repeated indices are summed)
\begin{equation}
	\mathcal{U}_{nk}\tilde{B}_k = 	 \mathcal{U}_{nk}\frac{\partial \tilde{A}_i}{\partial \tilde{x}_j}\epsilon_{ijk} =  \frac{\partial A_s}{\partial x_m} \mathcal{U}_{im}  \mathcal{U}_{js} \mathcal{U}_{nk}\epsilon_{ijk}  = B_n  {  .}
\end{equation}
Finally, the physical susceptibilities $\chi,\chi^{e}$ in the   anisotropic crystal as a function of applied fields $\bm{E},\bm{B}$ are
\begin{eqnarray}
	\label{aniso_trans}
	&& \bm{\chi}(\bm{E},\bm{B})=\frac{\partial^2 L}{\partial \bm{B}\partial \bm{B}}= \bm{\mathcal{U}}^{-1}[\bm{\tilde{\chi}}(\bm{\tilde{E}},\bm{\tilde{B}})]\bm{\mathcal{U}}^{-1} \nonumber\\
	&&  \bm{\chi^{e}}(\bm{E},\bm{B})=\frac{\partial^2 L}{\partial \bm{E}\partial \bm{E}}= \bm{\mathcal{U}}[\bm{\tilde{\chi}^{e}}(\bm{\tilde{E}},\bm{\tilde{B}})]\bm{\mathcal{U}} {  .}
\end{eqnarray}
The magnetization and polarization  vectors and higher order   susceptibility tensors transform in the usual way like above.
We re-iterate a point already mentioned in the main text that even when the crystal  anisotropy is not taken into account, that is $\bm{\mathcal{U}} =  \bm{1}$, the susceptibilities are intrinsically anisotropic, since they depend on the mutual alignment of electric and magnetic fields. Crystal anisotropy is an additional source of directional dependence  that is relevant in the experimental context. 

\subsection{Invariance of $E\cdot B$ and the transformation of the angle between $E$ and $B$}
\label{sec:parallel}
In account of anisotropy, the normalized fields are defined as Eq.~\eqref{eb}.
We note that the scalar product of externally applied fields  $\mathbf{E}\cdot \mathbf{B}$ transforms as
\begin{equation}
	\bm{E}\cdot \bm{B}= \mathcal{U}^{-1}_{ik}\tilde{E}_k \mathcal{U}_{il}\tilde{B}_l =   B_\star E_\star \mathbf{e}\cdot \mathbf{b},
\end{equation}
i.e., is proportional to the scalar product of the transformed fields. 
However, we note that $\bm{\mathcal{U}}$  is not a simple rotation. Hence, while scalar products are proportional, it does not preserve angles or norms taken separately. We see this when we express  $\mathbf{\hat{e}}\cdot \mathbf{\hat{b}}$ in terms of 
$\bm{\hat{E}}\cdot \bm{\hat{B}}$  as
\begin{equation}
	\label{angle_transform}
	\mathbf{\hat{e}}\cdot \mathbf{\hat{b}} = \frac{\bf{e}\cdot \bf{b}}{|\bf{e}| |\bf{b}|} =  \frac{\bm{E}\cdot \bm{B}}{|\bm{\mathcal{U}E}| |\bm{\mathcal{U}}^{-1}\bm{B}|}  = \frac{|\bm{E}| |\bm{B}| }{|\bm{\mathcal{U}E}| |\bm{\mathcal{U}}^{-1}\bm{B}|}\bm{\hat{E}}\cdot \bm{\hat{B}}
\end{equation}
Therefore
\begin{equation}
	\bm{E}\perp \bm{B} \iff \mathbf{e}\perp \mathbf{b}.
\end{equation}
We can work in a reference frame where $\bm{\mathcal{U}}$ is diagonal. However, even when $\bm{E}\parallel \bm{B}$ are directed along  a general direction, we have 
\begin{equation}
	\bm{E}\parallel \bm{B}\implies   \mathbf{\hat{e}}\cdot {\mathbf{\hat{b}}}= \frac{1}{|\bm{\mathcal{U}\hat{E}}| |\bm{\mathcal{U}}^{-1}\bm{\hat{E}}|} \leq 1
\end{equation}
The equality is satisfied when $\bm{E}$ and $\bm{B}$ are directed along the principle directions of the  frame that diagonalizes $\bm{\mathcal{U}}$, e.g.
$\bm{\hat{E}} = \bm{\hat{B}} = (1,0,0)^T$.  However, for general direction, the  vectors $\mathbf{e}$ and $\mathbf{b}$ are no longer parallel, for example, if
\begin{equation}
	\bm{\mathcal{U}} = \text{diag}(u,u,1/u^2),\quad \bm{\hat{E}} =\bm{\hat{B}}= \frac{1}{\sqrt{3}}(1,1,1)^T
\end{equation}
we obtain
\begin{equation}
	\mathbf{\hat{e}}\cdot {\mathbf{\hat{b}}}= \left(\frac{9}{{\left(\frac{2}{u^2 }+u^4 \right)}\,{\left(2\,u^2 +\frac{1}{u^4 }\right)}}\right)^{1/2}
\end{equation}
For exemplary values of $u =1.2,1.5,2$, we get the angle between $\mathbf{\hat{e}}$ and ${\mathbf{\hat{b}}}$ as $28^\circ,55^\circ,75^\circ$ respectively.

In terms of transformed vectors $\mathbf{e}$ and $\mathbf{b}$, Dirac materials are described by a Hamiltonian that has the usual isotropic form  which simplifies calculations. In particular, for the perpendicular fields $\mathbf{e}$ and $\mathbf{b}$, one can apply the Lorentz transformation that eliminates the smaller of the fields, resulting in the purely electric field or the purely magnetic field case. In the case when fields $\mathbf{e}$ and $\mathbf{b}$  are not orthogonal, a particular choice of a Lorentz transformation leads to a Hamiltonian with a new set of electric and magnetic fields parallel to each other, for which the calculations of dielectric and magnetic susceptibilities are simplified (Section~\ref{sec:asymptotic}).  Once the results are obtained in the frame where the  fields are parallel, the inverse Lorentz transformation allows to express them  in terms of the original non-orthogonal fields $\mathbf{e}$ and $\mathbf{b}$. Finally, inverse transformation to initial anisotropic system will give the results in terms of applied fields  $\bm{E}$ and  $\bm{B}$. 
In Sec.~\ref{sec:suscep} we derive the susceptibilities for an arbitrary angle between $\mathbf{e}$ and $\mathbf{b}$, see Table~\ref{tab:arbit} for quick reference.

\begin{figure}
	\centering
	\includegraphics[width=.6\columnwidth]{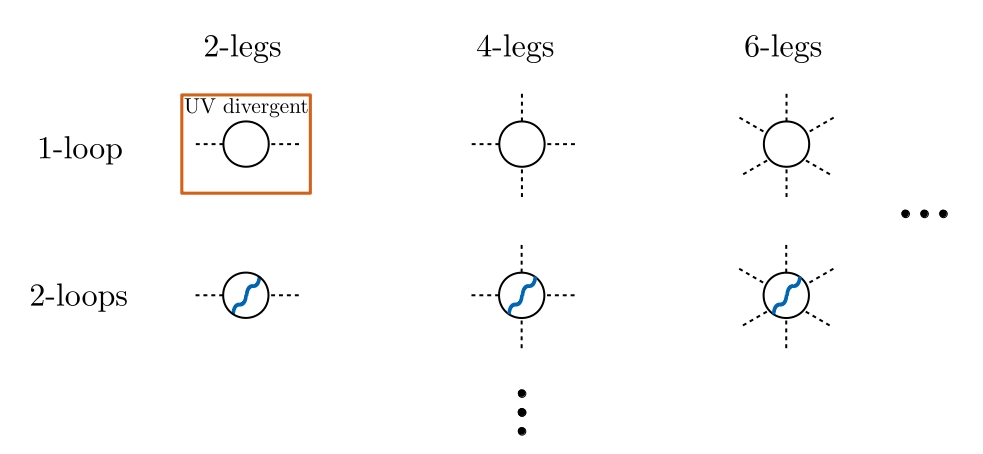}
	\caption{Multi-loop and multi-leg diagrams for the effective action. The dashed line corresponds to the  external electromagnetic fields $B$,$E$, the blue wavy lines are Coulomb interactions and the solid line is the Green function of a Dirac sea electron. Only the 1-loop, 2-leg diagram given by~\eqref{app:L1}, is ultraviolet divergent,  the others are convergent. The divergence is renormalized into the dielectric constant and relative permeability. The multi-loop diagrams contain Coulomb interaction (blue) that are suppressed by the relative permittivity, hence carry the factor $\alpha_D/\epsilon \sim 0.03$. The 1-loop diagrams with more than 4-legs are given by~\eqref{app:series}, when the expansion parameters $|\mathbf{e}|^2 =  \frac{\alpha_D E^2}{\Delta/\lambdabar_D^3}$  or $ |\mathbf{b}|^2 = \frac{\alpha_D B^2 v^2/c^2}{\Delta/\lambdabar_D^3}$, are large,  the diagrams can be exactly summed to yield Eq.~\eqref{app:Schwinger}}
	\label{fig:loop_diagrams}
\end{figure}

\begin{table*}[t]
	\centering
	\begin{tabular}{llll|llll}
		\hline\hline
		Parameter & \multicolumn{3}{l}{Dirac Insulator} & \multicolumn{3}{l}{QED}  \\
		\hline
		&symbol & expression &value$\sim$ & symbol& expression & value$\sim$ \\\hline
		Energy gap	& $2\Delta$& &$100\:\text{meV}$\footnote{This estimate is based on tunable gap compound $\text{Pb}_{1-x}\text{Sn}_x \text{Se}$~\cite{Dziawa2012}. There is no gap in Weyl, Dirac semimetals (see TaAs~\cite{Zhang2019}), and appropriate limit has to be taken. Gap is tunable between $\pm10\:\text{meV}$ in Bi and Sb alloys.  Typical gap for materials in Table~\ref{table:epsilons} is $200\:\text{meV}$. } & $2 m_e c^2$& &  $1\:\text{MeV}$  \\
		Dirac velocity	& $v$& $c/400$&  $10^6\:\text{m}/\text{s}$ & $c$ &&$ 3\times 10^8\:\text{m/s}$\\
		Wavelength (reduced) & $\lambdabar_D$& $ \frac{\hbar v}{\Delta}$&$ 10\:\text{nm}$ & $\lambdabar_C$&$ \frac{\hbar}{m_e c}$&$3.86\times 10^{-4}\:\text{nm} $\\
		Fine structure constant & $\alpha_D$&$ \frac{c\alpha}{v}$&$3$ & $\alpha$ &$ \frac{e^2}{\hbar c}$&$ 1/137 $\\
		Magnetic moment & $\mu_D$ &$ \frac{e \hbar v^2}{2\Delta c}$&$ 3.7\:\text{meV/T}$ & $\mu_B$& $ \frac{e\hbar}{2m_e c}$&$ 5.8\times10^{-2}\text{meV}/\text{T}$\\
		Ultraviolet cut-off	& $\Lambda$&& $ 2\:\text{eV}$ & $\Lambda_{QED}$&&$10^{28}\text{-}10^{286}\:\text{eV}$\footnote{Low estimate is based on Planck energy~\cite{birrell1984quantum}, where gravitational effects come into play. Large estimate is the Landau pole, where the coupling constant diverges.~\cite{landau_pole, *Suslov2009}} \\
		\thead[l]{Dielectric screening\\(due to Dirac sea)}	& $\epsilon_D$&$ 1+ \frac{2\alpha_D}{3\pi} \log\frac{\Lambda}{\Delta}$& $ 3$ & $\epsilon_\Lambda$&$1+ \frac{2\bar{e}^2}{3\pi \hbar c} \log\frac{\Lambda_{QED}}{m_e c^2}$~\footnote{See footnote g} \\
		\thead[l]{Relative permittivity\\
			(renormalized)} & $\epsilon $&$\sim\epsilon_D+ \epsilon_i$\footnote{$\epsilon_i$ is the lattice and intra-ionic contribution.}&$100$\footnote{See Table~\ref{table:epsilons}.} & $\epsilon$& & $1$\footnote{By definition, the vacuum has a permittivity of unity or $\epsilon_0$ in SI units. } \\
		Electric charge in use & bare: $e$&& $1.6 \times 10^{-19}\:\text{C}$ & RG: $e_0 = e$& $\bar{e}/\sqrt{\epsilon_\Lambda} $\footnote{The quantity $\bar{e}$ is the running charge defined at the ultraviolet cutoff energy.~\cite{Peskin2005}}&  $1.6 \times 10^{-19}\:\text{C}$\\
		Field definition in use & screened:  $E$ &$D/\epsilon$\footnote{$D$ is the macroscopic  electric displacement that is due to free charges only.} & & RG: $\bar{E} = E$&$  \sqrt{\epsilon_\Lambda} E_0$\\
		Schwinger E-Field & $E_\star$&$  \frac{\Delta^2}{e\hbar v}$&$ 5\times 10^4\:\text{V/cm}$ & $E_S$ &$ \frac{m_e^2 c^3 }{e \hbar}$&$ 1.3\times 10^{16} \:\text{V/cm}$ \\
		Schwinger B-Field & $B_\star $&$ \frac{\Delta^2 c}{e\hbar v^2}$&$ 6.8\:\text{T}$ & $B_S$&$  \frac{m_e^2 c^3 }{e \hbar}$&$ 4.4\times 10^{9} \:\text{T}$ \\
		\thead[l]{1-loop expansion \\
			parameter 1} & $|\bf{e}|^2$ &$\frac{\alpha_D E^2}{\Delta/\lambdabar_D^3}$&$<0.1$\footnote{Dielectric break down occurs beyond $|\bf{e}|\sim 0.1$ which is captured by our  nonperturbative theory, see Eq.~\eqref{app:breakdown}} & $|\bf{e}|^2$&$ \frac{  \alpha B^2 }{m_e c^2/\lambdabar_c^3}$&$<0.1$\footnote{Electric field accelerates free particle-hole pairs beyond this point.}\\
		\thead[l]{1-loop expansion \\
			parameter 2}  & $|\bf{b}|^2$ &$ \frac{\alpha_D B^2 }{\Delta/\lambdabar_D^3}\frac{v^2}{c^2}$& $< 10^3$\footnote{First Landau level is ejected beyond the cut-off energy past this point in external magnetic field. Therefore our theory breaks down. Note that perturbation theory is not valid beyond $|b|>1$, but the nonperturbative resummation in Eq.~\eqref{app:Schwinger} is still valid. } & $|\mathbf{b}|^2$&$ \frac{\alpha B^2 }{m_e c^2/\lambdabar_c^3}$& $<10^{44}$\footnote{Magnetic field energy curves space-time beyond this point.}\\
		\thead[l]{	N-loop expansion parameter\\
			(interaction probability)} & &$\alpha_D/\epsilon$&$0.03$\footnote{1-loop  nonperturbative theory is valid as long as $\alpha_D < \epsilon \sim 100$, that is $v \gtrsim 10^{-4}c $.} & $\alpha $& &$ \approx 0.01$ \\
		Invariant 1: $|\bf{e}|^2 - |\bf{b}|^2$  &$\propto\mathcal{F}^{\mu\nu}\mathcal{F}_{\mu\nu}$\footnote{$\mathcal{F}_{\mu\nu} = \tilde{\partial}_\mu \mathcal{A}_\nu - \tilde{\partial}_\nu \mathcal{A}_\mu$. This is different from true relativistic field tensor $F_{\mu\nu}= \partial_\mu A_\nu -\partial_\nu A_\mu$. See Eq.~\eqref{4-vectors}.} & $E^2 - \frac{v^2}{c^2}B^2$ &&$\propto{F}^{\mu\nu}{F}_{\mu\nu}$ & $E^2 - B^2$ \\
		Invariant 2: $\bf{e}\cdot\bf{b}$&$\propto\mathcal{F}_{\alpha\beta}\mathcal{F}_{\mu\nu}\varepsilon^{\alpha\beta\mu\nu}$&  $\frac{v}{c}\bm{E}\cdot\bm{B}$ &&$\propto{F}_{\alpha\beta}{F}_{\mu\nu}\varepsilon^{\alpha\beta\mu\nu}$ &$\bm{E}\cdot\bm{B}$
	\end{tabular}
	\caption{Comparison of  Dirac insulator with Quantum Electrodynamics (QED). For simplicity isotropic case is considered. For the  general case see Sec.~\ref{sec:aniso} }
	\label{table:compare}
\end{table*}

\section{Comparison to the relativistic quantum electrodynamics}
\label{sec:QED_compare}

The Dirac equation that follows from Eq.~\eqref{Dirac_iso}, after multiplying  from the left by $\bm{\beta}$, is
\begin{equation}
	i \bm{\gamma}^0(\partial_t -i|e|{\phi})     -  v  \bm{{\gamma}}\cdot  (\bm{\tilde{p}} + |e|\bm{\tilde{A}}/c)\psi-  \Delta \psi =0,\quad \bm{\gamma}^0 =\bm{\beta},\:\bm{\gamma}^i = \bm{\beta} \bm{\alpha}^i. 
\end{equation}
If we define the effective and true 4-position and electromagnetic 4-potential respectively as
\begin{align}
	\label{4-vectors}
	\text{effective:}&\quad \tilde{x}^\mu = (vt, \bm{\tilde{x}}),\quad \mathcal{A}^\mu = \frac{1}{v}(\phi, v\tilde{\bm{A}}/c)\\
	\text{true-relativistic:}&\quad	{x}^\mu = (ct, \bm{{x}}),\quad 	 {A}^\mu = \frac{1}{c}(\phi, \bm{A}),\quad \text{where } \eta^{\mu\nu} = (+,-,-,-),
\end{align}
we can write down the Dirac action coupled to classical electromagnetism in the (anisotropic) material as
\begin{equation}
	\label{Dirac_material_action}
	\text{Dirac material:}\quad S = \int dt d^3\tilde{x} \:\left( \bar{\psi}[ i v \gamma^\mu (\tilde{\partial}_\mu - i|e| \mathcal{A}_{\mu}) -\Delta]\psi +  \frac{1}{8\pi}( \bm{E}^T \bm{\epsilon} \bm{E} - \bm{B}^T \bm{\mu}^{-1} \bm{B})\right).
\end{equation}
In addition, when the band gap is inverted, as in topological insulators, there is an additional boundary term bulk-boundary term $\bm{E}\cdot \bm{B}$~\cite{Frohlich} which does not effect the nonlinear response which we investigate in this letter. For this reason we assume $\Delta>0$ with out loss of generality. 

For comparison, the  true relativistic QED action is
\begin{equation}
	\label{QED_action}
	\text{QED:}\quad S = \int dt d^3x \: \left(\bar{\psi}[ i c \gamma^\mu ({\partial}_\mu - i|e| {A}_{\mu}) -m_e c^2]\psi +  \frac{1}{8\pi}( {E}^2  - {B}^2)\right).
\end{equation}
The first difference one can notice is the speed of the Dirac fermion, that is $c$ in QED.  In the material  with a gap $2\Delta$, the Fermi-Dirac velocity is related to the effective mass $m^*$ (measured in terms of electron mass) and satisfies
\begin{align}
	\label{v_estimate}
	&E = \sqrt{\Delta^2 + p^2 v^2},\quad E \approx \Delta + \frac{v^2 p^2}{2 \Delta}\implies m^* m_e v^2 =\Delta,\nonumber\\ &\Delta \approx 0.1 \:\text{eV},\: m^* = 0.01\text{-}0.5\implies 1000\gtrsim c/v \gtrsim 100  { .}
\end{align}
For this reason, the effective fine structure constant of the insulator is
\begin{equation}
	\label{alpha_estimate}
	\alpha_D = \frac{e^2}{\hbar v} \sim \alpha \times 400 \approx 3  { .}
\end{equation} 
The minimally coupled 4-potentials are different as outlined in Eq.~\eqref{4-vectors}. The classical action of the electromagnetic field in the two cases are obviously different. In the material case,  $\bm{\epsilon},\bm{\mu}$ are symmetric matrices that define the relative permittivity and the relativity of the material, respectively. In the 
vacuum they satisfy $\epsilon=\mu = 1$ in the CGS units.

A more subtle difference between QED and the Dirac insulator is the definition of electric charge and electric field, when seen from a renormalization point of view.
In QED the charge is normalized at zero momentum transfer,
$e_0=e_{q=0}$ (referred to as the running electric charge~\cite{Peskin2005} ), which is related to the bare charge $\bar{e}$ through the ultraviolet (UV) cutoff dependent vacuum dielectric screening,  $e_0 = \bar{e}/\sqrt{\epsilon_\Lambda}$.  However, in condensed	matter, the charge is defined independently of scale and is given by $e \approx 1.6\times 10^{-19}\:\text{C}$. Nevertheless, due to RPA dielectric screening, the apparent charge at the macroscopic scale is $e/\sqrt{\epsilon}$, hence different from the bare $e$ that applies at the lattice spacing scale $q\sim \hbar\pi/a$.  Definitions of the
electric field are also different. In QED, the field is normalized to $\bm{\bar{E}} = \sqrt{\epsilon_\Lambda} \bm{E}$ so that the energy density of the field is fixed as $ \bar{E}^2/(8\pi)$, and the dielectric constant of vacuum is by definition equal to unity. However in condensed matter, the Lagrangian
of the free field, is already written in terms of the screened field $\bm{E}$, as $\epsilon E^2/(8\pi)$.
The coupling term is also written in terms of the bare charge and the screened field $e\bm{E}$, hence $\epsilon$ does not appear in the definition of the effective fine structure constant $\alpha_D = e^2/(\hbar v)$.
Importantly, the coupling term is independent of the definition,
$e_0{\bar {\bm E}}=e{\bm E}$, allowing us to use the 1-loop renormalized effective action of QED in the condensed matter setting.  A detailed comparison of parameters in Dirac insulator to their counterparts in QED is given in Table~\ref{table:compare}.

\section{Effective action for Dirac insulator}
\label{sec:eff_act}

The effective action can be \textit{schematically} organized in terms of multi-leg and multi-loop diagrams as in Fig.~\ref{fig:loop_diagrams}.
The multi-loop diagrams contain at least one internal interaction line and therefore in QED, each such line contains the factor $\alpha\sim 0.01$. 

We now justify the possibility to limit the calculation of the effective action to the 1-loop approach and justify neglecting multi-loop diagrams in Dirac materials, particularly insulators. In the Dirac insulator, we first  note that multi-loop diagrams are UV convergent hence the polarization is dictated mostly by the low energy-momentum sector where the static dielectric constant $\epsilon$ applies~\cite{Sham1966,Sham_Rice1966}. The  Coulomb interaction is therefore screened due to $\epsilon$, which is dominated by the lattice and intra-ionic polarization contribution.  In a wide range of practical situations as seen in Table~\ref{table:epsilons} $\epsilon \sim 100$ and therefore the interaction lines are suppressed by the factor $\frac{\alpha_D}{\epsilon}\sim 0.03$. 

We note that, no matter how small the interaction parameter $\alpha_D/\epsilon$, the multi-loop diagrams may come with large combinatorial factors and can render the perturbation series divergent. In both QED and condensed matter physics several examples of divergences are known. In some of these divergences, the coupling constant $\alpha$ gets enhanced by an additional parameter $R$, $\alpha\rightarrow R\alpha$. For example, in an electron gas in metals, there is Random Phase Approximation (RPA) infrared divergence, where $\alpha$ is enhanced by a large distance. This diverging series of diagrams is eventually converging, and can be re-summed, as was shown by Gell-Mann Brueckner.~\cite{Gell-Mann_RPA} However, in the case of Dirac insulator, the electron gas infrared divergence does not arise. Indeed, for example the expansion of contribution to energy due to interactions,
\begin{equation}
	\label{RPA}
	E= E_0+ \langle 0|H_{ee}|0\rangle + \sum_n \frac{\langle 0|H_{ee}|n\rangle \langle n|H_{ee}|0\rangle} {E_n-E_0}+....,
\end{equation}
where $E_0$ is the interaction-independent contribution to energy, $H_{ee}$ is the interaction Hamiltonian, index zero denotes the ground state of the system and index n corresponds to excited states. In an electron gas, the denominator 
$E_n-E_0=\frac{\hbar^2}{m}(q^2 + ({\bm k}_1-{\bm k}_2)\cdot {\bm q})$, where ${\bm k}_1$ and ${\bm k}_2$ are momenta (wave vectors) of the two holes inside the Fermi sphere, ${\bm q}$ is the transferred momentum,  ${\bm k}_2+{\bm q}$ and ${\bm k}_2-{\bm q}$ are the momenta of electrons outside the Fermi sphere. This denominator is divergent in the second order perturbation theory, and higher order terms are even more divergent. For the electron gas, these perturbation series can be summed. However, for the Dirac insulator, when excitations arise only in transitions between the full valence band and empty conduction band, every denominator in series Eq. (\ref{RPA}) contains extra constant $2\Delta$, and divergences do not appear. Furthermore, as we consider high magnetic fields, if both spin states of the highest occupied Landau level are fully filled,
excitation involves a cyclotron energy gap. Then if the cyclotron energy is much larger than the characteristic Coulomb interaction, the mixing of excited states is negligible, and the non-interacting "closed shell 
configuration" is essentially the ground state~\cite{giuliani2005quantum}. Due to large $\epsilon$, this happens in rather small magnetic fields in our case. Such situation is relevant for our consideration of Bi, where we consider the limit $\Delta=0$ and zero electric field. It becomes clear that electron gas RPA-like contribution need not be included in our consideration.

Another type of divergence often arising due to interactions 
is Lippman-Schwinger~\cite{Lippmann_Schwinger}, Bethe-Salpeter effects~\cite{Bethe_Salpeter}, or, in condensed matter context, the divergence related to the exciton bound state.~\cite{Wannier,Mott,Schro} For exciton, when $\epsilon$ is big, binding energy is small and the size is effectively big.  Small binding energy means closeness in energy to the bottom of the conduction band. For this bound state to manifest itself in susceptibilities, excitations have to be generated at frequencies of electromagnetic wave close to $2\Delta$. Thus, the related contributions can be separated from the effects at small frequencies that we mostly consider.

In susceptibilities, divergences as a result of summation of series of interaction contributions may also indicate phase transitions, such as ferromagnetism. In the presence of the gap $2\Delta$ for excitations in Dirac insulators, such phase transition is not expected. Similarly, a superconducting transition, while emerging due to binding of electrons in Cooper pairs associated with  the presence of the Fermi surface (see, e.g. Abrikosov~\cite{Abrikosov}), is unlikely to arise in the presence of excitation gap in insulators. 

We note that the one-loop Heisenberg-Euler contribution at large B that we calculated, Eq.~\ref{strong} of the main text, also stems from divergence of the general renormalization group type, when $\alpha$ is enhanced by the first power of log. 

Finally, there is potentially a divergence stemming from loop expansion that differ from the cases discussed above. It contains the fine structure constant $\alpha$ only, and is asymptotic, with combinatorial coefficients growing fast with increasing the power $n$ of the fine structure constant in diagrams with $n$ interaction lines and vertices of electromagnetic interactions. These so-called asymptotic series were considered first by Dyson~\cite{Dyson} (Also see  an excellent review by Huet and co-authors~\cite{multi_loop_qed}). It is now widely accepted that for asymptotic series a summation up to an infinite order in small $\alpha$ does not make sense, and one must terminate the series  after a few terms when doing practical calculations with QED, although a rigorous mathematical justification for using 1-loop QED is an open problem.

Thus, we can safely restrict ourselves to 1-loop approximation since
\begin{equation}
	\label{multi_loop_parameter}
	\frac{(K+1)\text{-loops}, 2n\text{-legs} }{K\text{-loops}, 2n\text{-legs}}\sim \frac{\alpha_D}{\epsilon} \approx 0.03,\quad \text{See Eq.~\eqref{alpha_estimate} and Table~\ref{table:epsilons}}	  { .}
\end{equation}

This applies to QED where $\alpha \approx 1/137$, as considered by Heisenberg and Euler, and in the condensed matter contexts, when the interaction parameter $e^2/(\hbar v \epsilon)$ is small at large $\epsilon$, fixed by lattice properties. We note that our predictions pass the test of comparison with available experimental results (See Fig.~\ref{fig:epsilon_M} of the main text), even in the strong field regime which can not be directly probed in QED. 

\begin{table*}[t]
	\centering
	\begin{tabular}{llll}
		Compound       & $2\Delta$[eV]    & $\epsilon$ & Refs.   \\ 
		\hline
		\ce{PbTe}    &   $0.19$      & $450$  & \cite{Lin1966,burstein,Svane2010,Kanai_1963}    \\
		\ce{PbSe}    &   $0.16$       & $280$ & \cite{Lin1966,burstein,Svane2010}   \\
		\ce{PbS}  &   $0.28$  &$190$    &\cite{Lin1966,burstein,Svane2010}  \\
		\ce{SnTe}  &   $0.27$      & $1770$    &\cite{Suzuki_1995,Hermann_1968,Hayasaka_2016} \\
		\ce{Sb2Te3}  &   $0.21$     & $168,36$    &\cite{Lefebvre_1987, Springer_AntimonyTelluride_epsilon}  
		\\
		\ce{Bi2Te3}  &   $0.17$        & $290,75$    &\cite{Ando2013,Springer_BismuthTelluride_epsilon}    
		\\
		\ce{Bi2Se3}  &   $0.30$        & $113$    &\cite{Ando2013,Springer_BismuthSelenide_epsilon}  \\
		GeBi$_{4\text{-}x}$Sb$_x$Te$_7$  &   $0.10$        & $448,30$~\footnote{@ $x=0$, i.e. \ce{GeBi4Te7}}    &\cite{Ando2013,Jain2013}  
		\\
		\ce{GeBi2Te4}  &   $0.18$       & $315,19$    &\cite{Ando2013,Jain2013}  
		\\
		$\text{Bi}_{1-x}\text{Sb}_{x}$ &   $0.01$~\footnote{gap vanishes at $x=0.04$}       & $\sim 100$~\footnote{@ x = 0.07, where $2\Delta \approx 0.01\:\text{eV}$}    &\cite{ion_polarization,Rudolph_bisb}  	  
	\end{tabular}
	\caption{The  Dirac gap and dielectric constants of Dirac materials. When multiple numbers for $\epsilon$ {  are} provided,   anisotropy is signified  (see references for more information). The linear dielectric contribution from the valence bands coming from Eq.~\eqref{app:L1}  is, by definition, already contained in these measurements.}
	\label{table:epsilons}
\end{table*}

\section{1-Loop effective action}
\label{sec:1-loop}
The 1-loop effective action of the material follows as
\begin{equation}
	\label{app:1loopformal}
	S_{eff} = -i \ln \mathrm{Det}\left[	 v \gamma^\mu (\tilde{\partial}_\mu - i|e| \mathcal{A}_{\mu}) +i \Delta\right]   { .}
\end{equation}
This determinant is exactly calculated in the QED case~\cite{Dunne,*Dunne_arxiv}. By comparing the QED and Dirac insulator actions in Eqs.~\eqref{Dirac_material_action} and~\eqref{QED_action},  we can write the 1-loop,  nonperturbative, renormalized Heisenberg-Euler action in terms of the normalized fields defined in Eqs.~\eqref{eb} and \eqref{eb_crit} as~\cite{Akhiezer1965,Berestetskii1982}
\begin{eqnarray}	
	\label{app:Schwinger}
	\delta L_{HE} &=&
	\frac{\Delta}{8\pi^2\lambdabar_D^3} \int_0^\infty \frac{d\eta e^{-\eta}}{\eta^3}\left[(-\eta A \cot (\eta A) \eta C \coth(\eta C)
	+ 1 - \frac{1}{3} \eta^2 (A^2-C^2)\right],\nonumber\\
	A&=&-\frac{i}{2}\left[\sqrt{({\bf b}+i{\bf e})^2}
	-\sqrt{({\bf b}-i{\bf e})^2}\right],  \nonumber\\
	C&=&\frac{1}{2}\left[\sqrt{({\bf b}+i{\bf e})^2}
	+\sqrt{({\bf b}-i{\bf e})^2}\right]  { }.
\end{eqnarray} 
In addition to this, there is also the cut-off dependent part of the action that is due to the UV-divergent diagram (shown in Fig.~\ref{fig:loop_diagrams} in a red box) which evaluates to
\begin{equation}
	\label{app:L1}
	\delta L_1  = 
	\frac{\Delta }{12\pi^2\lambdabar_D^3}\ln\left(\frac{\Lambda}{\Delta}\right)\left( |\mathbf{e}|^2 -  |\mathbf{b}|^2\right).
\end{equation}
In QED, this quantity has the same form as the classical electromagnetic Lagrangian{ ,} hence it is renormalized into the definition of electric charge. However in the material case, it gives rise to the magnetic and electric susceptibilities that we discuss below absorbed into $\bm{\epsilon},\bm{\mu}$ of the material. 
The Eq.~\eqref{app:Schwinger} is valid for both strong and weak fields. The contour on the complex $\eta$ plane shall be chosen so that the poles in the integrand are avoided. The imaginary part of the action gives the breakdown probability (volume rate of particle hole creation in QED)  { . For example, } when $|\mathbf{b}|=0$, we have to leading order~\cite{Berestetskii1982}
\begin{equation}	
	\label{app:breakdown}
	P =2\text{Im}\:\delta L_{HE} = \frac{\Delta}{4\pi^3\hbar\lambdabar_D^3}|{\bf e}|^2e^{-\pi/|{\bf e}|},
\end{equation}


Meanwhile, another way to write Eq.~\eqref{app:1loopformal} is the standard form
\begin{equation}
	S_{eff} = -i \ln \mathrm{Det}\left[	 v \gamma^\mu (\tilde{\partial}_\mu - i|e| \mathcal{A}_{\mu}) +i \Delta\right] = -i\text{Tr}\ln [1 +  |e|   v G \gamma^\mu  \mathcal{A}_{\mu}] - i \ln \mathrm{Det}\left[-i G^{-1}\right]  { .}
\end{equation}
Carrying out the formal diagrammatic expansion of the logarithm{ ,} we have
\begin{equation}
	1\text{-loop},2n\text{-legs}   =  \frac{i}{2 n} \text{Tr}[(|e|   v  \gamma^\mu  \mathcal{A}_{\mu} G)^{2n}], \quad n = 1,2,... 
\end{equation}
For example, the 2-leg diagram after taking  the Fourier transform $\mathcal{A}(q)= \int d^4\tilde{x}  \mathcal{A}(\tilde{x}) e^{i k^\mu \tilde{x}_\mu}$ reads
\begin{equation}
	1\text{-loop},2\text{-legs}   = 	\frac{i e^2 v^2}{4}\int \frac{d^4 q d^4 k}{(2\pi)^8 }   \text{tr}\left[ \slashed{\mathcal{A}}(q) \frac{1}{v\slashed{k}-\Delta} \slashed{\mathcal{A}}(-q)\frac{1}{v(\slashed{k}+\slashed{q})-\Delta}\right],\quad k^\mu = (\omega/v, \bm{\tilde{k}}),\: \slashed{k} = \gamma^\mu k_\mu  { ,}
\end{equation}
which gives the UV divergent term Eq.~\eqref{app:L1}.
The higher order diagrams can be identified by formally expanding the action Eq.~\eqref{app:Schwinger} in the fields. For example{ ,} if we choose $\mathbf{e}\parallel \mathbf{B}$, we have~\cite{Dunne,*Dunne_arxiv}
\begin{equation}
	\label{app:series}
	1\text{-loop},2n\geq 4\text{-legs} = -\frac{\Delta}{8\pi^3 \lambdabar_D^3} (2n-3)! \sum_{k=0}^n \frac{\mathcal{B}_{2k}\mathcal{B}_{2n-2k}}{(2k)!(2n-2k)!} (2 |\mathbf{b}|)^{2n-2k} (2i |\mathbf{e}|)^{2k}  { ,}
\end{equation}
where $\mathcal{B}_{2n}$ are Bernoulli numbers. The expansion parameters are
\begin{equation}
	|\mathbf{e}|^2 = \alpha_D E^2 \frac{\lambdabar_D^3}{\Delta},\quad 	|\mathbf{b}|^2 = \alpha_D B^2 \frac{v^2}{c^2}\frac{\lambdabar_D^3}{\Delta}  { .}
\end{equation}
Therefore when the fields or the coupling  constant are strong, the diagrammatic expansion is only schematic because the series Eq.~\eqref{app:series} should be summed exactly back into the original form Eq.~\eqref{app:Schwinger}, where the non-coplanar $\bm{E}$ and $\bm{B}$ case is also taken into account.

\section{Regime of validity of 1-loop effective action for a Dirac material}
\label{sec:validity}

From previous sections, firstly, we have to ensure that the 1-loop approximation holds by keeping the interaction strength under control
\begin{equation}
	\alpha_D \ll \epsilon \sim 100,\quad \text{(1-loop approximation holds)}  { .}
\end{equation}
When the magnetic field is too large, the   Landau level separation energy becomes comparable to the cut-off, and the Dirac description breaks down,  but at
\begin{equation}
	|\bm{B}| \ll \frac{\Lambda^2 c}{e \hbar v^2}, 
\end{equation}
the Dirac theory is valid. Finally when the E-field is strong, a formal asymptotic expression can be easily derived from Eq.~\eqref{app:Schwinger}. However, in our case this becomes a somewhat an academic question,  because as seen from Eq.~\eqref{app:breakdown} the electric breakdown probability increases significantly with strong fields, which we do not consider.
At small fields we have conditions
\begin{align}
	|\bm{E}|&\ll \frac{\Lambda^2}{e \hbar v}, \quad \text{(The Dirac theory is valid)}\\
	|\mathbf{e}| &< 0.3, \quad \text{(avoids  electric breakdown)}.
\end{align}

\section{Asymptotic expansion of the Heisenberg-Euler Action  }
\label{sec:asymptotic}

In order to avoid electric breakdown of an insulator, we are always in the weak $E$ limit, where the parallel field Lagrangian, being quadratic in $|\mathbf{e}|^2$ and $\hat{\bf{e}}\cdot \hat{\mathbf{b}}$,  can be expanded as

\begin{table}[]
	\begin{tabular}{llll}
		\multicolumn{1}{l|}{$f_{n}^m(b,x)$ } &       $n\rightarrow$               &                       &  \\ \hline\hline
		\multicolumn{1}{l|}{$m\downarrow$} & \multicolumn{1}{l|}{$b^2 \left(\frac{1}{x^2} + \frac{1}{3} - \frac{\coth(x)}{x}\right)$} & \multicolumn{1}{l|}{$-\left(1 + \frac{x}{2b}\right) \left(\frac{1}{x^2} + \frac{1}{3} - \frac{\coth(x)}{x}\right)$} & $\frac{x(x+b)}{8 b^4} \left(\frac{1}{x^2} + \frac{1}{3} - \frac{\coth(x)}{x}\right)$  \\ \cline{2-4} 
		& \multicolumn{1}{l|}{} & \multicolumn{1}{l|}{$\frac{1}{2\sinh^2(x)} - \frac{3-2x^2}{6x}\coth(x)$} & $\left(\frac{1}{b^2}-\frac{x}{2b^3}\right)\left(\frac{1}{2\sinh^2(x)} - \frac{3-2x^2}{6x}\coth(x)\right)$ \\ \cline{3-4} 
		&                       & \multicolumn{1}{l|}{} & $-\frac{6 x\coth(x) + 4 x^2 + 9 }{24 b^2\sinh^2(x)} + \coth(x)\frac{225-60x^2 + 8x^4}{360 b^2 x}$ \\ \cline{4-4} 
	\end{tabular}  { .}
	\caption{The functions $f_n^m(x,b)$ appearing in the integrands in $I^m_n(b)$ (Eq.~\eqref{app:expansion_integrals})  the perturbation expansion of the renormalized Lagrangian Eq.~\eqref{app:weak_a_strong_b_expansion}}
	\label{table:fs}
\end{table}

\begin{table}[]
	\begin{tabular}{llll}
		\multicolumn{1}{l|}{$I_{n}^m(b)$ } &       $n\rightarrow$               &                       &  \\ \hline\hline
		\multicolumn{1}{l|}{$m\downarrow$} & \multicolumn{1}{l|}{$ \ln(b) \left(\frac{1}{3}b^2 + b + \frac{1}{2}\right)  + b  + \frac{3}{4} + \frac{\pi^2}{72 b} - \frac{\zeta(3)}{48b^2}$} & \multicolumn{1}{l|}{$-\ln(b)\left(\frac{1}{3} + \frac{1}{2b} \right) - \frac{1}{6} - \frac{1}{b} - \frac{1}{4b^2}$ } & $\frac{1}{12 b^2}$  \\ \cline{2-4} 
		& \multicolumn{1}{l|}{} & \multicolumn{1}{l|}{$\frac{b}{3} + \frac{1}{2b}\left( \ln(b) + 1\right) + \frac{\zeta(3) + 3}{12 b^2}$} & $\frac{1}{6b}$ \\ \cline{3-4} 
		&                       & \multicolumn{1}{l|}{} & 0 \\ \cline{4-4} 
	\end{tabular}  { .}
	\caption{Strong $B$-field expansion (up to order $1/b^2$) of the integrals $I^m_n(b)$ (Eq.~\eqref{app:expansion_integrals}) in the perturbation expansion of the renormalized Lagrangian Eq.~\eqref{app:weak_a_strong_b_expansion}}
	\label{table:Is}
\end{table}

\begin{equation}
	\label{app:weak_a_strong_b_expansion}
	\delta L_{HE} = \frac{\Delta}{8\pi^2\lambdabar_D^3}\sum_{n=0}^{\infty} \sum_{m=0}^n |\mathbf{e}|^{2n} (\mathbf{\hat{e}}\cdot\mathbf{\hat{b}})^{2m} I_{n}^{m}(b),\quad  b= |\mathbf{b}|  { .}
\end{equation}
The functions $I_n^m$ are integrals of the form
\begin{equation}
	\label{app:expansion_integrals}
	I_n^m(b) = \int_0^\infty dx \frac{ e^{-x/b}}{x} f_n^m(b,x)  { .}
\end{equation} 
where the functions $f_n^m(b,x)$ are tabulated in Table~\ref{table:fs}. 

Making a change of variables $x\to b x'$, we can 
recheck the low field limit of the Lagrangian in Eq.~\eqref{weak} of the main text
\begin{equation}
	\frac{8\pi^2\lambdabar_D^3}{\Delta}\delta L_{HE} \to \frac{1}{45}\int_0^{\infty} dx\: xe^{-x}\left(b^4  + 7 (\mathbf{e}\cdot\mathbf{b})^2 -|\mathbf{e}|^2 b^2 \frac{x+2}{2}  + |\mathbf{e}|^4 \frac{x + x^2}{8} \right) = \frac{1}{45} [(|\mathbf{e}|^2-|\mathbf{b}|^2) + 7(\mathbf{e}\cdot\mathbf{b})]  { .}
\end{equation}

The strong $B$ limit is obtained if we use the fact that for a bounded function $g(x)\to 0$ when $x\to 0$  {  the} integral converges to
\begin{equation}
	\label{app:asymptote}
	\int_0^{\infty} \frac{dx}{x} e^{-x\delta} g(x) \to g(\infty) \ln(1/\delta),\quad \delta \to 0  { .}
\end{equation}
Taking {  the} derivative or integral with respect to $1/b$, we can generate the subleading or super-logarithmic terms{ ,} respectively. Performing this procedure we obtain the strong $B$ expansions of the integrals in Eqs.~\eqref{app:weak_a_strong_b_expansion} and \eqref{app:expansion_integrals}, as tabulated in Table~\ref{table:Is}. Reading of the leading order terms from Table~\ref{table:Is} and substituting in Eq.~\eqref{app:weak_a_strong_b_expansion}  we obtain Eq.~\eqref{strong} of the main text
\begin{equation}
	\frac{8\pi^2\lambdabar_D^3}{\Delta}\delta L_{HE} \to \frac{1}{3} |\mathbf{b}|^2 \log |\mathbf{b}| + \frac{1}{3}|\mathbf{e}|^2 |\mathbf{b}|(\mathbf{\hat{e}}\cdot\mathbf{\hat{b}})^2.
\end{equation}


\begin{figure}
	\includegraphics[width=
	0.4\textwidth]{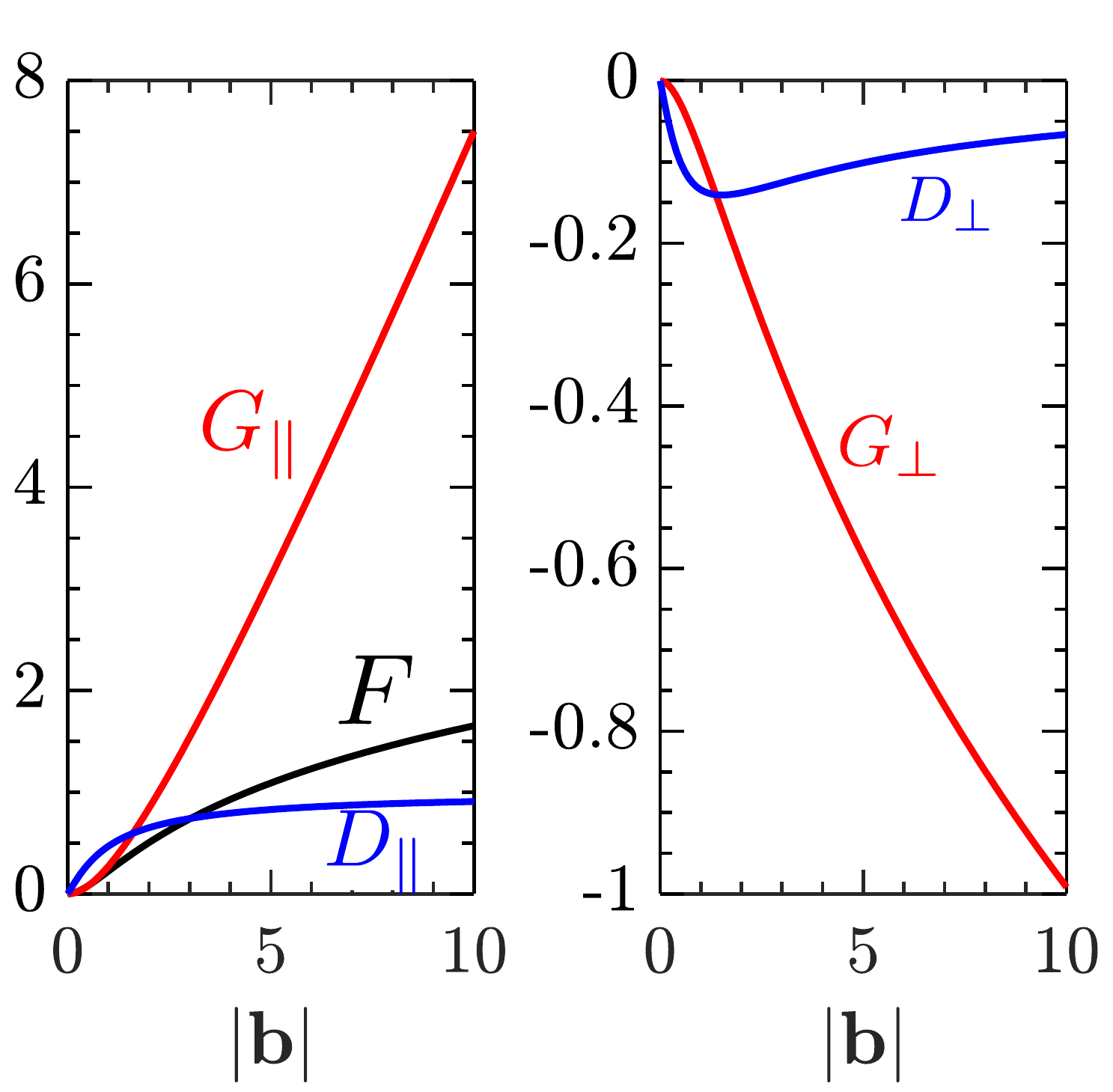}
	\caption{Dimensionless functions F,G,D of Eqs.~\eqref{dchi4}, \eqref{deps}) and \eqref{magn} versus the dimensionless magnetic field. Functions G and D depend on the relative orientation of electric and magnetic field. The left (right)  panel corresponds to 	$\mathbf{b} \parallel \mathbf{e}$ ($\mathbf{b}  \perp  \mathbf{e}$). For arbitrary angle between $\mathbf{b}, \mathbf{e}$ see Table~\ref{tab:arbit}.}
	\label{fig:FGD}
\end{figure}

\begin{table}[]
	\begin{tabular}{l|l|l|l}
		
		\multirow{2}{*}{$\bm{\delta \chi}$} & \multirow{2}{*}{$\frac{v^2}{c^2}\frac{\alpha_D}{12\pi^2}\bm{\mathcal{U}}^{-2}$ } & $2|\mathbf{b}|^2/5$ & $|\mathbf{b}|\ll 1$ \\ 
		&                   &$\log(|\mathbf{b}|)$  & $|\mathbf{b}|\gg 1$ \\ \hline
		\multirow{2}{*}{$\bm{\delta\epsilon(B)}$} & \multirow{2}{*}{$\frac{\alpha_D}{3\pi}\bm{\mathcal{U}}^{2}$} & $\frac{7}{15}(\mathbf{\hat{e}}\cdot\mathbf{\hat{b}})^2 - \frac{2}{15} |\mathbf{b}|^2$  & $|\mathbf{b}|\ll 1$ \\ 
		&                   & $-\log(|\mathbf{b}|) + (\mathbf{\hat{e}}\cdot\mathbf{\hat{b}})^2 |\mathbf{b}|$ & $|\mathbf{b}|\gg 1$ \\ \hline
		\multirow{2}{*}{$4\pi \bm{M^{(e)}}$}         & \multirow{2}{*}{$\frac{\mu_D}{3\pi \lambdabar_D^3 }|\mathbf{e}|^2 \bm{\mathcal{U}}^{-1}$} & $ \frac{2}{15}\left(   7\mathbf{\hat{e}} (\mathbf{\hat{e}}\cdot \mathbf{b})-2 \mathbf{b}\right)$ & $|\mathbf{b}|\ll 1$ \\ 
		&                   & $-\hat{\mathbf{b}}(\mathbf{{\mathbf{\hat{e}}}}\cdot \mathbf{\hat{b}})^2 + 2{\mathbf{\hat{e}}}(\mathbf{{\hat{e}}}\cdot \mathbf{\hat{b}}) -\frac{\mathbf{\hat{b}}}{|\mathbf{b}|}$ & $|\mathbf{b}|\gg 1$  \\
	\end{tabular}
	\caption{  Asymptotic expressions for the nonlinear magnetic susceptibility $\bm{\chi}$, magnetically modulated dielectric constant $\bm{\delta \epsilon (B)}$ and the electric modulated magnetization $4\pi \bm{M^{(e)}}$ tensors for arbitrary angle between $\mathbf{\hat{e}}$ and $\mathbf{\hat{b}}$.}
	\label{tab:arbit}
\end{table}

\section{Derivation of Susceptibility tensors}
\label{sec:suscep}
Inspecting the classical Lagrangian of the electromagnetic field $L_0 = (E^2 -  B^2)/(8\pi^2)$, we can derive the (linear {  and} nonlinear) contributions to susceptibilities from Eq.~\eqref{app:L1} and Eq.~\eqref{app:Schwinger}.  If the quantum part of the action is called $\delta L$,  the differential magnetic and electric susceptibilities are
\begin{subequations}
	\label{app:Chi}
	\begin{align}
		\chi_{ij} &=   \frac{\partial^2 \delta L}{\partial B_j ial B_i} = \frac{v^2\alpha_D \lambdabar_D^3}{c^2\Delta} \frac{\partial^2 \delta L}{\partial \text{b}_l \partial \text{b}_k} \mathcal{U}^{-1}_{ik}\mathcal{U}^{-1}_{jl}   { ,}\\
		\chi_{ij}^{e} &=   \frac{\partial^2 \delta L}{\partial {E}_j \partial {E}_i}  =\frac{\alpha_D \lambdabar_D^3}{\Delta} \frac{\partial^2 \delta L}{\partial \text{e}_l \partial \text{e}_k}  \mathcal{U}_{ik}\mathcal{U}_{jl} { .}
	\end{align}
\end{subequations}
%

\subsection{Linear and low order nonlinear susceptibilities}
\label{sec:low_order}

The cutoff dependent UV contribution is
\begin{equation}
	\delta L_1  = 
	\frac{\Delta }{12\pi^2\lambdabar_D^3}\ln\left(\frac{\Lambda}{\Delta}\right)\left( |\mathbf{e}|^2 -  |\mathbf{b}|^2\right).
\end{equation}
In vacuum quantum electrodynamics{ ,} this quantity has the same form as the classical electromagnetic Lagrangian{,} hence it is absorbed into the definition of electric charge. However in the material case, it gives rise to the isotropic valence{  band} contribution to{  the magnetic and electric susceptibilities,} respectively 
\begin{equation}
	\label{app:Chi_Gamma}
	(\tilde{\chi}_1)_{ij} = -\frac{v^2}{c^2} \frac{\alpha_D}{6\pi^2} \ln\left(\frac{\Lambda}{\Delta}\right)\delta_{ij} = 
	-\frac{v^2}{c^2}(\tilde{\chi}^{e}_1)_{ij}  { .}
\end{equation}
In the anisotropic case, these become
\begin{equation}
	\bm{\chi_D} = -\frac{v^2}{c^2} \frac{\alpha_D}{6\pi^2} \ln\left(\frac{\Lambda}{\Delta}\right)\bm{\mathcal{U}}^{-2},\quad \bm{\epsilon_D} = \bm{1}+	4\pi \bm{\chi_D^{e}} = \bm{1} +  \frac{2\alpha_D}{3\pi^2} \ln\left(\frac{\Lambda}{\Delta}\right)\bm{\mathcal{U}}^{2}  { .}
\end{equation}
as in Eqs.~\eqref{ec} and \eqref{ec2} of the main text.

The weak field limit of Eq.~\eqref{app:Schwinger} contains the lowest order nonlinear{  terms that are }quadratic in $|\mathbf{e}|^2-|\mathbf{b}|^2$ and $\mathbf{e}\cdot \mathbf{b}$ 
\begin{equation}
	\label{app:lag_2}
	\delta L_2=
	\frac{\Delta}{360\pi^2\lambdabar_D^3}
	\left[\left(|\mathbf{e}|^2- |\mathbf{b}|^2\right)^2
	+7\left({\bf e}\cdot{\bf b}\right)^2\right],
\end{equation} 
from which the nonlinear differential susceptibility tensors for an isotropic system follow as
\begin{subequations}
	\label{supp:weak}
	\begin{align}
		\delta\tilde{\chi}_{ij} &\to \frac{v^2}{c^2}\frac{\alpha_D }{180\pi^2}\left[2|\mathbf{b}|^2 (2\hat{b}_i \hat{b}_j + \delta_{ij}) + |\mathbf{e}|^2 (7\hat{e}_i\hat{e}_j - 2\delta_{ij})\right],\\
		\delta\tilde{\epsilon}_{ij} &\to \frac{\alpha_D }{45\pi}\left[ |\mathbf{b}|^2 (7\hat{b}_i\hat{b}_j - 2\delta_{ij}) + 2|\mathbf{e}|^2 (2\hat{e}_i \hat{e}_j + \delta_{ij}) \right], \quad \hat{e}_i = \frac{\text{e}_i}{|\mathbf{e}|},\: \hat{b}_i = \frac{\text{b}_i}{|\mathbf{b}|}
	\end{align}  
\end{subequations}
These quantities can be re-expressed in the anisotropic case by using Eq.~\eqref{aniso_trans} as
\begin{subequations}
	\label{supp:weak_aniso}
	\begin{align}
		\bm{\delta\chi} &=\frac{v^2}{c^2} \frac{\alpha_D}{180\pi^2} \left(  4 (\bm{\mathcal{U}^{-1}}\mathbf{b}) \otimes (\bm{\mathcal{U}^{-1}}\mathbf{b})+  2 |\mathbf{b}|^2\bm{\mathcal{U}}^{-2}  + 7(\bm{\mathcal{U}^{-1}}\mathbf{e}) \otimes (\bm{\mathcal{U}^{-1}}\mathbf{e})-2\bm{\mathcal{U}}^{-2}|\mathbf{e}|^2\right)\label{chi_weak},\\
		\bm{\delta\epsilon} &=\frac{\alpha_D}{45\pi} \left(  4 (\bm{\mathcal{U}}\mathbf{e}) \otimes (\bm{\mathcal{U}}\mathbf{e})+  2 |\mathbf{e}|^2\bm{\mathcal{U}}^{2}  + 7  (\bm{\mathcal{U}}\mathbf{b}) \otimes (\bm{\mathcal{U}}\mathbf{b})-2\bm{\mathcal{U}}^{2}|\mathbf{b}|^2\right).\label{epsilon_weak}
	\end{align}  
\end{subequations}
The electric modulated magnetization then is simply $\bm{M^{(e)}} = \bm{\delta\chi(E) B} = B_\star \bm{\delta\chi(E) \mathcal{U} }\mathbf{b}$ and reads
\begin{equation}
	\label{M_weak}
	4\pi \bm{M^{(e)}}  = \frac{2\mu_D}{45\pi \lambdabar_D^3}\bm{\mathcal{U}}^{-1}\left(   7\mathbf{e} (\mathbf{e}\cdot \mathbf{b})-2 \mathbf{b}|\mathbf{e}|^2\right).
\end{equation}

To obtain the expressions in the main text we work in coordinates where
\begin{equation}
	\bm{\mathcal{U} }= \text{diag}(U_{xx}, U_{yy}, U_{zz}).
\end{equation}
For simplicity we assume $\bm{E}$ and $\bm{B}$ directed along principal directions, e.g. $\bm{E}\parallel \bm{B}\parallel \mathbf{\hat{x}}$ for the parallel field configuration and  $\bm{E}\parallel  \mathbf{\hat{x}},\:\: \bm{B}\parallel\mathbf{\hat{y}}$ for the  perpendicular field configuration.  When these conditions are met, we can reduce, for example the parallel weak field magnetic susceptibility  as
\begin{equation}
	(\bm{\delta\chi_{\parallel} (E)} )_{xx} =\bm{\mathcal{U}}^{-2}_{xx}\frac{v^2}{c^2} \frac{\alpha_D}{36\pi^2} |\mathbf{e}|^2, \quad 	4\pi(\bm{M^{(e)}_{\parallel}})_{x} = (\bm{\mathcal{U}}^{-1} \mathbf{b})_{x}  \frac{2\mu_D}{9\pi \lambdabar_D^3}|\mathbf{e}|^2.
\end{equation}
For a summary of susceptibilities for arbitrary mutual orientation of E\&B see Table~\ref{tab:arbit}.
As we discussed in section S1, using inverse anisotropy transformation, the results for susceptibilities can be re-written in terms of applied external fields $\bm{E}$ and $\bm{B}$. The angle between $\mathbf{e}$ and $\mathbf{b}$ as function of the angle between $\bm{E}$ and $\bm{B}$ is given in Eq.~\eqref{angle_transform}.

\subsection{Higher nonlinear corrections to susceptibilities}

Now that we have the renormalized Lagrangian in the form Eq.~\eqref{app:weak_a_strong_b_expansion}, with the integrals $I^m_n$ tabulates in Table~\ref{table:Is}, we can calculate the susceptibilities according to Eq.~\eqref{app:Chi}, in notable cases. If the electric field is zero{ ,} we have  
\begin{equation}
	\frac{8\pi^2\lambdabar_D^3}{\Delta} \delta L_{HE}(|\mathbf{e}| = 0) \to \frac{\mathbf{b}^2}{3}\ln(\mathbf{b}) + \mathbf{b}\ln(\mathbf{b})+...
\end{equation} 
consistent with Eq.~\eqref{strong} of the main text.
Therefore the contribution to magnetic susceptibility due to applied magnetic field is
\begin{equation}
	\label{chi_strong}
	\bm{\delta\chi}(|\mathbf{e}|=0) \to  \frac{v^2}{c^2}\frac{\alpha_D}{12\pi^2}\bm{\mathcal{U}}^{-2}\ln |\mathbf{b}|   { ,}
\end{equation}
as in Eq.~\eqref{dchi4} of the main text. 

The term in the Lagrangian that is second order in the electric field   { is}
\begin{equation}
	\frac{8\pi^2\lambdabar_D^3}{\Delta} \frac{\partial \delta L_{HE}}{\partial |\mathbf{e}|^2}\bigg|_{|\mathbf{e}|=0} = -\ln (|\mathbf{b}|) \left(\frac{1}{3} + \frac{1}{2|\mathbf{b}|}\right) + (\mathbf{\hat{\mathbf{e}}}\cdot\mathbf{\hat{b}})^2\left(\frac{|\mathbf{b}|}{3} + \frac{1}{2|\mathbf{b}|}\left(\ln |\mathbf{b}| +1\right) 
	\right) + ...
\end{equation}
Then the leading order magnetization contribution is obtained from $M = \partial L/\partial B$ as
\begin{equation}
	\label{M_strong}
	4\pi \bm{ M}^{(\bm{e})} \to \frac{\mu_D}{3\pi\lambdabar_D^3}|\mathbf{e}|^2 \bm{\mathcal{U}^{-1}}\left(-\hat{\mathbf{b}}(\mathbf{{\mathbf{\hat{e}}}}\cdot \mathbf{\hat{b}})^2 + 2{\mathbf{\hat{e}}}(\mathbf{{\hat{e}}}\cdot \mathbf{\hat{b}}) -\frac{\mathbf{\hat{b}}}{|\mathbf{b}|}...\right),
\end{equation}
as in Eq.~\eqref{magn} of the main text. 
%
The dielectric response, linear in the electric field, is due to the term in the Lagrangian that is second order in the electric field{ ,} from which we obtain the magnetic field contribution in the isotropic case as
\begin{equation}
	\label{epsilon_strong}
	\bm{\delta{\epsilon}} \to \frac{\alpha_D}{3\pi} \left(-\ln (|\mathbf{b}|)(\bm{\mathcal{U}}\mathbf{\hat{e}})\otimes (\bm{\mathcal{U}}\mathbf{\hat{e}}) + (\bm{\mathcal{U}}\mathbf{\hat{b}})\otimes (\bm{\mathcal{U}}\mathbf{\hat{b}}) |\mathbf{b}| \right)  { .}
\end{equation}

The leading order electric field contribution to the dielectric tensor comes from the term that is $\sim |\mathbf{e}|^4$ in the Lagrangian

\begin{equation}
	\delta{\tilde{\epsilon}}_{ij} \to \frac{\alpha_D}{6\pi}\frac{|\mathbf{e}|^2}{\mathbf{|b|}^2}\left(\delta_{ij} +2 \hat{e}_i \hat{e}_j +  |\mathbf{b}| \delta_{ij}(\mathbf{\hat{e}}\cdot \mathbf{\hat{b}})^2 + 2|\mathbf{b}|[\hat{e}_i \hat{b}_j + \hat{e}_j \hat{b}_i](\mathbf{\hat{e}}\cdot \mathbf{\hat{b}}) + \mathbf{|b|}\hat{e}_i \hat{e_j}\right)  { ,}
\end{equation}
hence small in the parameter $|\mathbf{e}|^2/\mathbf{|b|}\ll 1$.

%
%
%
%

\section{Material Applications}
\label{sec:applications}

\subsection{Bismuth and $\text{Bi}_{0.9}\text{Sb}_{0.1}$	}
\label{Bi_estimates}

While our consideration is primarily for insulators  , we, e.g., demonstrated that the calculated magnetic susceptibility would be valid for valence band contributions in gapless materials. However real experiments often include, particularly for insulators with small gap, a certain density of free electrons. Then analysis of experimental settings must take into account contributions from free electrons, or experimental conditions must be found when these contributions are suppressed.

Bismuth has a band gap of $2\Delta = 15.5\:\text{meV}$ at the $L$-point and a Fermi level of $\mu = 35\:\text{meV}$~\cite{Fuseya2015} measured from the midgap point of the Dirac bands. The electronic Fermi surface is composed of 3 electron ellipsoids that lie on the binary-bisectrix  (x-y) plane perpendicular to the trigonal (z) axis. There is also a hole pocket along the trigonal axis. The hole Fermi level is about $-196\:\text{meV}$ measured from the midgap point of the hole Dirac bands.  The hole band gap is about $2\Delta = 370\:\text{meV}$.~\cite{Liu1995} 

The hole contribution to susceptibility in the binary-bisectrix plane is small $\chi \sim -10^{-7}$ due to the large hole band gap and the alignment of the hole ellipsoid.

The conduction bands are polarized { in magnetic field, so that only the e1 ellipsoid is populated when a field is applied in the binary direction~\cite{Iwasa2019}. Furthermore, the e1 conduction electrons} can exhibit de  Haas-van Alphen oscillations, which are suppressed above $B=5\:\text{T}$. At higher magnetic fields the nonlinear diamagnetism is identical to the insulating alloy $\text{Bi}_{0.9}\text{Sb}_{0.1}$, which is identical to bismuth except for the absence of the Fermi level.

Based on the band structure calculations~\cite{Liu1995}, we write the diagonalized velocity tensor as
\begin{equation}
	\bm{\mathcal{V}} = \text{diag}(1.7, \:1.5,\: 0.4) v ,\quad c/v = 188,
\end{equation}
The Dirac cone is located at the L-point, therefore the velocity operator in the crystal frame is obtained by using an appropriate rotation-reflection operator.
The critical field is about $B_\star = 36\:\text{mT}$. 
We find the total susceptibility by adding the contributions from each L-point located symmetrically in the binary axis
\begin{equation}
	\bm{\chi_{1}}+ \bm{\delta \chi}(\bm{B}) = \sum_{n=1}^3 (\bm{R}^T)^{n} \left[\bm{\chi_{1}}+\bm{\delta \chi}\left(\frac{\bm{\mathcal{U}}^{-1} \bm{R}^{n} \bm{B}}{B_\star}\right)\right]^{(L-point\:n)} \bm{R}^{n-1},
\end{equation}
where $R$ implements rotation by $120^\circ$ about the trigonal axis. 
The first order contribution is
\begin{equation}
	\bm{\chi_D} =-10^{-5}\begin{pmatrix}
		3 & 0 & 0\\
		0 & 3 & 0\\
		0 & 0 & 0.5
	\end{pmatrix}.
\end{equation}
If we apply magnetic field in the binary ($x$) direction we have
\begin{equation}
	\delta\chi_{xx}(B_x = 5\:\text{T}) = 13 \times 10^{-6}.
\end{equation}
For the full field range see Fig.~\ref{fig:epsilon_M}. 
The other quantities (dielectric enhancement, magnetization etc.) are calculated in a similar fashion.
For example the dielectric enhancement when $E \parallel B$ is  in $x$-direction we have
\begin{equation}
	\delta\epsilon_\parallel(B_x) \to 10 B_x.
\end{equation}
Given that the dielectric constant of the bismuth alloy is about $\epsilon~100$ (See Table~\ref{table:epsilons}) arising mostly from the ionic crystal, the magnetic field induced nonlinear contribution due to the Dirac band is enormous even at relatively weak B-fields.

To compare the nonlinear behavior of our theory with the experiment by Iwasa \textit{et al.}~\cite{Iwasa2019}, we take the reference level of susceptibility to be $\chi_{ref} = -\chi(50\:\text{T})$.

According to our theory 
\begin{equation}
	\chi_{theory}^{binary}(50\:\text{T}) = -5.5\times 10^{-6},\quad \chi_{theory}^{bisectrix}(50\:\text{T}) = -8.0\times 10^{-6}.
\end{equation}
The Iwasa experiment has the reference points
\begin{equation}
	\chi_{exp}^{binary}(50\:\text{T}) = -18.7\times 10^{-6},\quad \chi_{exp}^{bisectrix}(50\:\text{T}) = -15.4\times 10^{-6}.
\end{equation}
The difference in the reference levels are due to the additional contributions to linear magnetic susceptibility that are not contained in the Dirac theory. Although the logarithmic Dirac contribution in Eq.~\eqref{ec2} constitutes a significant part of the linear diamagnetic response,   non-Dirac core shell electron contributions { can be} equally important. Furthermore, there can be additional errors due to the choice of UV cut-off and subleading terms in the summation of the Landau levels can contribute to the total. None of the above mentioned factors affect the nonlinear response, which is dominated by the Dirac bands. Therefore, once the overall constant offset due to linear susceptibility is adjusted, we get an excellent agreement in the nonlinear behavior of the measured susceptibility in bismuth, as seen in Fig.~\ref{fig:chis}.

\subsection{TaAs}
\label{TaAs_estimates}
TaAs has 12 pairs of Weyl fermions~\cite{Weyl_cones}, 4 pairs forming hole pockets $2\:\text{meV}$ below the Fermi level and 8 forming electron pockets $21\:\text{meV}$ above the Fermi level. The Fermi level $E_F = 21\:\text{meV}$ contribution which is not accounted in our theory, becomes unimportant $B \gg \frac{4 E_F^2}{v_1 v_2 e\hbar } \approx 10\:T.$

We represent each pair by a single massless Dirac fermion with the velocity tensor comparable to the calculations and measurements~\cite{Zhang2019,Weyl_cones}
\begin{equation}
	\bm{\mathcal{V}} = \text{diag}(1.7, \:1.7,\: 0.35) v ,\quad c/v = 447
\end{equation}
where the the z-component is the velocity in the $c$-axis.
Since the system is gapless we consider the situation where $E = 0$. The nonlinear susceptibility is the gapless limit is
\begin{equation}
	\bm{\chi_D +\delta\chi} \to -\frac{\alpha_D}{12\pi^2} \frac{v^2}{c^2} \ln\left(\frac{c\Lambda^2}{e B \hbar v^2}\right)\bm{\mathcal{U}}^{-2}
\end{equation}

%
%
%

To compare the nonlinear behavior of our theory with the experiment by Zhang \textit{et al.}~\cite{Zhang2019}, we take the reference level of susceptibility to be $\chi_{ref} = -\chi(30\:\text{T}) = 1.56\times 10^{-5}$.

%


\subsection{$\text{Pb}_{0.8}\text{Sn}_{0.2}\text{Te}$}
\label{Lead_estimates}
The gap as a function of the Sn content is~\cite{Hayasaka_2016}
\begin{equation}
	2\Delta[\text{meV}] = 182 - 480x  
\end{equation}
At $x = 0.235$ we have the inverse masses $m^{-1}_\perp  = 100 m^{-1}_e$ and $m^{-1}_\parallel = 11.25 m_e^{-1}$ where the gap is $2\Delta = 69.1\:\text{meV}$.
At $x = 0.510$ we have the same inverse massed  where the gap is similar $2\Delta = -62.8\:\text{meV}$ (TI phase). We will base our estimates on these values	. The parallel denotes the z-axis aligned in the [001] direction.
The velocity tensor is then
\begin{equation}
	\bm{\mathcal{V}} = \text{diag}(1.4, \:1.4,\: 0.5) v ,\quad c/v = 580.
\end{equation}

There are a total of 4 Dirac fermions located at the L-points symmetric about the z-axis [001] direction of the rock salt structure. Due to the relatively large gap the most striking property is the electric field modulated magnetization.
If we apply $\bm{B}\parallel \bm{E}$  in the $x$-direction with $B\sim 5\:T$ and $E = E_\star/3$ we have
\begin{equation}
	4\pi ({M^{e}_\parallel})_x(B_x = 5\:\text{T}, E_x = 10^4\:\text{V/cm}) \approx 0.5\:\mu \text{T}.
\end{equation}

\end{document}